\documentclass[12pt,a4paper]{article}
\usepackage[british]{babel}
\usepackage{epsfig}
\begin{document}
\date{}

\title{
{\vspace{-20mm} \normalsize
\hfill \parbox[t]{50mm}{\small DESY 05-250     \\
                               MS-TP-05-33     \\
                               SFB/CPP-05-83}} \\[15mm]
 Numerical simulations with two flavours of \\
 twisted-mass Wilson quarks and DBW2 gauge action\\[5mm]}

\author{F.\ Farchioni$^{a}$,
        P.\ Hofmann$^{a}$,
        K.\ Jansen$^{b}$,
        I.\ Montvay$^{c}$,\\
        G.\ M\"unster$^{a}$,
        E.E.\ Scholz$^{c}$\footnote{Present address: Physics Department,
         Brookhaven National Laboratory, Upton, NY 11973 USA},
        L.\ Scorzato$^{d}$,
        A.\ Shindler$^{b}$,\\
        N.\ Ukita$^{c}$\footnote{Present address: Institute of Physics,
         University of Tsukuba, Tsukuba, Ibaraki 305-8571, Japan and 
         Department of Physics, The University of Tokyo,
         Hongo 7-3-1, Bunkyo-ku, Tokyo 113-0033, Japan},
        C.\ Urbach$^{b,e}$\footnote{Present address:
        Theoretical Physics Division, Dept. of Mathematical Sciences, University of
        Liverpool, Liverpool L69 3BX, UK},
        U.\ Wenger$^{b}$\footnote{Present address:
        Institute for Theoretical Physics, ETH Z\"urich, CH-8093 Z\"urich, Switzerland},
        I.\ Wetzorke$^{b}$\\[5mm]
  {\small $^a$ Universit\"at M\"unster,
   Institut f\"ur Theoretische Physik,}\\
   {\small Wilhelm-Klemm-Strasse 9, D-48149 M\"unster,
   Germany}\\
  {\small $^b$ NIC/DESY Zeuthen, Platanenallee 6, D-15738 Zeuthen,
   Germany}\\
  {\small $^c$ Deutsches Elektronen-Synchrotron DESY, Notkestr.\,85,
   D-22603 Hamburg, Germany}\\
  {\small $^d$ Institut f\"ur Physik, Humboldt Universit\"at zu Berlin,
   D-12489 Berlin, Germany}\\
  {\small $^e$ Freie Universit\"at Berlin,
   Institut f\"ur Theoretische Physik,}\\
  {\small Arnimallee 14, D-14196 Berlin, Germany}\\[5mm]}
%
\newcommand{\be}{\begin{equation}}                                              
\newcommand{\ee}{\end{equation}}                                                
\newcommand{\half}{\frac{1}{2}}                                                 
\newcommand{\rar}{\rightarrow}                                                  
\newcommand{\lar}{\leftarrow}
\newcommand{\LCB}{\raisebox{-0.3ex}{\mbox{\LARGE$\left\{\right.$}}}
\newcommand{\RCB}{\raisebox{-0.3ex}{\mbox{\LARGE$\left.\right\}$}}}
\newcommand{\LSB}{\raisebox{-0.3ex}{\mbox{\LARGE$\left[\right.$}}}
\newcommand{\RSB}{\raisebox{-0.3ex}{\mbox{\LARGE$\left.\right]$}}}
\newcommand{\tr}{{\rm Tr}}

\newcommand{\amcp}{am_\chi^{\rm\scriptscriptstyle PCAC}}
\newcommand{\mcp}{m_\chi^{\rm\scriptscriptstyle PCAC}}
\newcommand{\mcpr}{m_{\chi R}^{\rm\scriptscriptstyle PCAC}}

\newcommand{\amcpb}{a\bar{m}_\chi^{\rm\scriptscriptstyle PCAC}}
\newcommand{\mcpb}{\bar{m}_\chi^{\rm\scriptscriptstyle PCAC}}

\newcommand{\amqp}{am_q^{\rm\scriptscriptstyle PCAC}}
\newcommand{\mqp}{m_q^{\rm\scriptscriptstyle PCAC}}

\newcommand{\cpp}{\chi'_{\rm\scriptscriptstyle PCAC}}

 
\maketitle

\abstract{
 Discretisation errors in two-flavour lattice QCD with Wilson-quarks and
 DBW2 gauge action are investigated by comparing numerical simulation
 data at two values of the bare gauge coupling.
 Both non-zero and zero twisted mass values are considered.
 The results, including also data from simulations using the Wilson
 plaquette gauge action, are compared to next-to-leading order chiral
 perturbation theory formulas.}

\newpage
\section{Introduction}\label{sec1}

 The singular point of QCD at vanishing quark masses is distorted in
 Wilson-type lattice formulations: as a result of lattice artefacts, in
 the region of small quark masses an extended phase structure is
 developed.
 This phase structure can be predicted and classified in Chiral
 Perturbation Theory (ChPT) \cite{CHPT} if lattice artefacts are taken
 into account \cite{SHARPE-SINGLETON}.
 If, in addition to the usual quark mass parameter, a twisted quark
 mass is introduced \cite{AOKI-PHASE,TMQCD} then in the plane of
 untwisted and twisted quark mass a first order phase transition
 {\em line} with second order endpoints appears.
 Depending on the sign of the leading term representing lattice
 artefacts, the first order phase transition line is either on the
 untwisted quark mass axis (``Aoki-phase scenario'' \cite{AOKI-PHASE})
 or perpendicular to it (``normal scenario'')
 \cite{MUENSTER,SCORZATO,SHARPE-WU1}.

 In numerical simulations it pays off to try to reduce lattice
 artefacts at fixed (non-vanishing) lattice spacing by an appropriate
 choice of the lattice action.
 An important issue in this respect is to bring the phase structure at
 small quark masses as close as possible to the point-like singularity
 appearing in the continuum limit.
 In fact, the strong first order phase transition observed earlier in
 numerical simulations with Wilson-type quarks
 \cite{BLUMETAL,JLQCD,JANSEN} presents a serious obstacle for QCD
 simulations with light quarks.

 In previous work we systematically investigated the phase structure of
 lattice QCD with twisted-mass Wilson-type quarks (for a recent review
 see \cite{SHINDLER}).
 In Ref.~\cite{TWIST} we have shown that at lattice spacings near
 $a \simeq 0.2{\rm\,fm}$ the phase structure with Wilson-quarks and
 Wilson-plaquette gauge action is consistent with the ``normal
 scenario'' of ChPT.
 This differs from the situation in the strong coupling regime, where
 the ``Aoki-phase scenario'' has been previously observed
 \cite{HUMBOLDT}.

 A consequence of the ``normal scenario'' is that for fixed gauge
 coupling ($\beta$) the mass of charged pions have a positive lower
 bound ($m_\pi^{min}$).
 The numerical simulation data in Ref.~\cite{TWIST} have shown that this
 lower bound is at $a \simeq 0.2{\rm\,fm}$ quite high, namely about
 $600{\rm\,MeV}$.
 Such a high lower bound would prohibit the study of light quarks.
 Therefore, an important question is the behaviour of this lower
 bound as a function of the gauge coupling (or lattice spacing) towards
 the continuum limit.
 In a subsequent paper it has been shown \cite{PLAQUETTE} that, as
 expected, the lower bound becomes clearly smaller for decreasing
 lattice spacing.
 Its decrease in the range $0.20{\rm\,fm} \geq a \geq 0.14{\rm\,fm}$ is
 roughly consistent with the prediction of next-to-leading-order (NLO)
 ChPT \cite{SHARPE-SINGLETON,MUENSTER,SCORZATO,SHARPE-WU1,SHARPE-WU2},
 namely $m_\pi^{min} \propto a$ (at $a\mu=0$).
 A minimal pion mass of $m_\pi^{min} \simeq 300{\rm\,MeV}$ is estimated
 to occur near $a \approx 0.07-0.10{\rm\,fm}$, but this estimate is
 rather uncertain and has to be checked in future simulations if the
 Wilson gauge action ought to be used.
 The question arises whether one could lower $m_\pi^{min}$ by a
 suitable change of the lattice action.

 An early observation by the JLQCD Collaboration has been \cite{JLQCD}
 that the strength of the first order phase transition near zero quark
 mass is sensitive to a change of the gauge action. 
 Following this hint, we have shown in a previous paper \cite{DBW} that
 combining two flavours ($N_f=2$) of Wilson-quarks with the DBW2 gauge
 action \cite{DBW2-ACTION} leads to a phase structure near zero quark
 mass with substantially weaker first order phase transition.
 As a consequence, the minimal pion mass is at least by a factor of two
 lower compared to the plaquette gauge action at similar lattice
 spacings.

 This implies that numerical simulations with light quarks become
 possible on coarser lattices and hence with much less computational
 costs if the DBW2 gauge action is used.
 Of course, for the choice of the gauge action also other criteria may
 be relevant.
 For instance, it has been reported in quenched studies
 \cite{NECCO,TAKEDA} that in some quantities strong scale breaking
 effects appear if the DBW2 action is used.
 Another problem could be the late convergence of lattice perturbation
 theory, implied by the results of the QCDSF Collaboration
 \cite{QCDSF}.

 In general, the question of the scaling behaviour of the results
 obtained by a given lattice action is very important.
 In case of the Wilson twisted-mass formulation of lattice QCD it has
 been shown \cite{FREZZOTTI-ROSSI} that the leading lattice artefacts
 are of ${\cal O}(a^2)$ if the bare quark masses are appropriately
 tuned.
 Detailed investigations have shown \cite{ZEUTHEN1,REGINA,ZEUTHEN2} that
 in the quenched approximation excellent scaling behaviour can be
 achieved, indeed, also at light quark masses.
 The same question in the full theory with dynamical quarks is
 obviously very important.

 In the present paper we perform first exploratory scaling tests for
 the combination of Wilson-fermion lattice action with the DBW2 gauge
 action by comparing numerical simulation data at two values of the
 gauge coupling, namely $\beta=0.67$ and $\beta=0.74$.
 We consider data points with both vanishing and non-vanishing value of
 the twisted mass.
 Moreover, since one can extract useful information on multiplicative
 renormalisation factors from the dependence of matrix elements on the
 twist angle in the plane of untwisted and twisted quark mass,
 we exploit this method and derive from our simulation data the values
 of $Z_V$, $Z_A$ and $Z_P/Z_S$.
 In addition, we compare the NLO-ChPT formulas of
 Refs.~\cite{MUENSTER,SCORZATO,SHARPE-WU1,SHARPE-WU2,MUENSTER-SCHMIDT}
 to the results of the numerical simulations.
 For comparison, ChPT fits of the data obtained by the Wilson plaquette
 gauge action \cite{PLAQUETTE} are also considered.

 The outline of the paper is as follows: in the next section, after
 specifying the lattice action and the simulation algorithms,
 the numerical simulation runs are discussed and some scaling tests are
 presented.
 Section \ref{sec3} is devoted to a detailed description of the results
 on the twist angle in the plane of untwisted and twisted quark mass
 together with an explanation how the aforementioned multiplicative
 renormalisation $Z$-factors can be determined.
 The knowledge of the twist angle and $Z$-factors makes it possible to
 obtain results on physical quantities, such as the quark mass and the
 pion decay constant.
 In Section \ref{sec4} the ChPT fits of the data with DBW2 gauge action
 are presented.
 Section \ref{sec5} contains a discussion and a summary.
 In an Appendix alternative chiral fits of the DBW2 data are shown
 and compared to similar ChPT fits of Wilson plaquette data.

\section{Numerical simulations}\label{sec2}

 The lattice action and simulation algorithms are defined here for the
 reader's convenience.
 The notations are similar to those in Ref.~\cite{DBW}.

\subsection{Lattice action and simulation algorithms}\label{sec2.1}

 We apply for quarks the lattice action of Wilson fermions, which can be
 written as
\be\label{eq01}
S_q = \sum_x \left\{ 
\left( \overline{\chi}_x [\mu_\kappa + i\gamma_5\tau_3\,a\mu ]\chi_x \right)
- \half\sum_{\mu=\pm 1}^{\pm 4}
\left( \overline{\chi}_{x+\hat{\mu}}U_{x\mu}[r+\gamma_\mu]\chi_x \right)
\right\} \ .
\ee
 Here the (``untwisted'') bare quark mass in lattice units is
 denoted by
\be\label{eq02}
\mu_\kappa \equiv am_0 + 4r = \frac{1}{2\kappa} \ ,
\ee
 $r$ is the Wilson-parameter, set in our simulations to $r=1$, $am_0$
 is another convention for the bare quark mass in lattice units and
 $\kappa$ is the conventional hopping parameter.
 The twisted mass in lattice units is denoted here by $a\mu$.
 (This differs from the notation in \cite{DBW} where $\mu$ has been
 defined without the lattice spacing factor $a$ in front.)
 $U_{x\mu} \in {\rm SU(3)}$ is the gauge link variable and we also
 defined $U_{x,-\mu} = U_{x-\hat{\mu},\mu}^\dagger$ and
 $\gamma_{-\mu}=-\gamma_\mu$.

 For the SU(3) Yang-Mills gauge field we apply the DBW2 lattice action
 \cite{DBW2-ACTION} which belongs to a one-parameter family of actions
 obtained by renormalisation group considerations.
 Those actions also include, besides the usual $(1\times 1)$ Wilson loop
 plaquette term, planar rectangular $(1\times 2)$ Wilson loops:
\be\label{eq03}
S_g = \beta\sum_{x}\left(c_{0}\sum_{\mu<\nu;\,\mu,\nu=1}^4
\left\{1-\frac{1}{3}\,{\rm Re\,} U_{x\mu\nu}^{1\times 1}\right\}
+c_{1}\sum_{\mu\ne\nu;\,\mu,\nu=1}^4
\left\{1-\frac{1}{3}\,{\rm Re\,} U_{x\mu\nu}^{1\times 2}\right\}
\right) \ ,
\ee
 with the condition $c_{0}=1-8c_{1}$.
 For the DBW2 action we have $c_1=-1.4088$.

 For preparing the sequences of gauge configurations two different
 updating algorithms were used: the Hybrid Monte Carlo (HMC) algorithm
 \cite{HMC} with multiple time scale integration and mass
 preconditioning as described in \cite{HMC-UJSW} and the two-step
 multi-boson (TSMB) algorithm \cite{TSMB} which has been tuned for QCD
 applications following \cite{NF2TEST,TWIST}.

\subsection{Simulation parameters and a first scaling test}\label{sec2.2}

 In our numerical simulations we considered two values of the gauge
 coupling, namely $\beta=0.67$ and $\beta=0.74$.
 The simulations at the lower $\beta$-value have been performed on a
 $12^3 \cdot 24$ lattice as in \cite{DBW}.
 The higher $\beta$-value ($\beta=0.74$) was chosen in such a way that
 the physical volume of the $16^3 \cdot 32$ lattice remains
 approximately the same, that is
 $a(\beta=0.74) \simeq \frac{3}{4}\,a(\beta=0.67)$.
 The value of the lattice spacing was defined by extrapolating the
 Sommer scale parameter in lattice units $r_0/a$ \cite{SOMMER} to zero
 quark mass and assuming $r_0 \equiv 0.5{\rm\,fm}$.
 The simulation parameters and the amount of statistics are specified in
 Table \ref{tab_run}.

 As Table \ref{tab_run} shows, both zero and non-zero twisted mass
 points were simulated.
 The non-zero values of the twisted mass were also chosen according to
 the assumed scale ratio, that is
 $a\mu(\beta=0.74)=\frac{3}{4}\,a\mu(\beta=0.67)=0.0075$.
 In other words, the bare twisted mass $\mu$ is kept (approximately)
 constant.

 In several points of the parameter space simulation runs have been
 performed with both the HMC and the TSMB updating algorithms.
 Having run the two algorithms in the same points allowed to compare
 their performance.
 It turned out that the optimised HMC algorithm of Ref.~\cite{HMC-UJSW}
 is substantially faster than TSMB.
 For instance, in long runs at the simulation point ($A$)
 ($16^3 \cdot 32$ lattice, $\beta=0.74$, $\kappa=0.1580$, $a\mu=0$) HMC
 with multiple time scale integration and mass preconditioning is almost
 by a factor of 10 faster.
 Therefore, in the majority of simulation points the final data analysis
 is based on HMC runs.
 Results from TSMB updating were only used in the runs of the first
 part of Table \ref{tab_run} (those at $\beta=0.67$ and $a\mu=0$).
 Even if results with both updating algorithms were available in several
 other points, in the final analysis we never mixed results from 
 different updating procedures.

 The results for some basic quantities are collected in Tables
 \ref{tab_one} and \ref{tab_two}.
 The pseudoscalar meson (``pion'') mass $am_\pi$ is obtained from the
 correlator of the charged pseudoscalar density
\be\label{eq04}
P^{\pm}_x=\bar\chi_x\frac{\tau^\pm}{2}\gamma_5\chi_x
\ee
 where $\tau_\pm \equiv \tau_1 \pm i\tau_2$.
 In case of the vector meson (``$\rho$-meson'') mass $am_\rho$, for
 generic values of the bare untwisted and twisted quark mass, the
 correlators of both vector ($V^a_{x\mu}$) and axialvector
 ($A^a_{x\mu}$) bilinears of the $\chi$-fields can be used:
\be\label{eq05}
V^a_{x\mu} \equiv \overline{\chi}_x \half\tau_a\gamma_\mu \chi_x \ ,
\hspace{3em}
A^a_{x\mu} \equiv \overline{\chi}_x \half\tau_a\gamma_\mu\gamma_5
\chi_x
\hspace{3em}
(a=1,2) \ .
\ee
 The reason is that the physical vector current is, in general, a
 linear combination of $V^a_{x\mu}$ and $A^a_{x\mu}$ (see Section
 \ref{sec3}).
 In a given simulation point we determined $am_\rho$ from the
 correlator possessing the better signal.

 In Table~\ref{tab_two} the values of the bare (untwisted) PCAC quark
 mass $\amcp$ are also given.
 It is defined by the PCAC-relation containing the axialvector current
 $A^a_{x\mu}$ in (\ref{eq05}) and the pseudoscalar density
 $P^{\pm}_x$:
\be\label{eq06}
\amcp \equiv
\frac{\langle \partial^\ast_\mu A^+_{x\mu}\, P^-_y \rangle}
{2\langle P^+_x\, P^-_y \rangle} \ .
\ee
 Here $\partial^\ast_\mu$ denotes, as usual, the backward lattice
 derivative.

 Besides $\amcp$, Table~\ref{tab_two}
 also contains the values of the bare ``untwisted'' pseudoscalar decay
 constant $af_{\chi\pi}$ defined by
\be\label{eq07}
af_{\chi\pi} \equiv (am_\pi)^{-1} 
\langle 0 | A^+_{x=0,0} | \pi^- \rangle \ .
\ee
 The relation of the bare (untwisted) quantities
 $\amcp$ and $af_{\chi\pi}$ to the
 corresponding physical quantities will be discussed in the following
 section.

\begin{figure}
\begin{center}
{\Large\bf Figures}
\end{center}
\vspace*{1em}
\begin{center}
\vspace*{0.10\vsize}
\begin{minipage}[c]{1.0\linewidth}
\hspace{0.10\hsize}
\includegraphics[angle=-90,width=0.75\hsize]
 {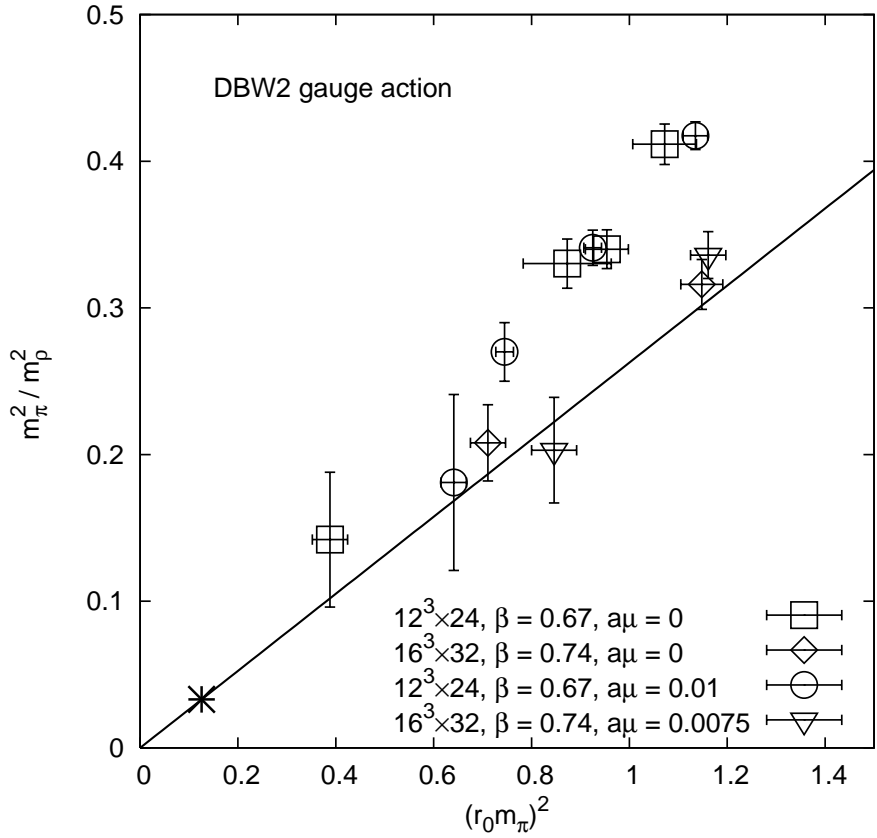}
\end{minipage}
\end{center}
\vspace*{-2em}
\begin{center}
\parbox{0.8\linewidth}{\caption{\label{fig_mPimRho}\em
 The squared pion to $\rho$-meson mass ratio $(m_\pi/m_\rho)^2$ versus
 $(r_0 m_\pi)^2$.
 Only simulation points with positive quark mass are considered.
 The physical point is shown by an asterisk.
 The straight line connecting the origin with it is the continuum
 expectation for small quark masses where both quantities are
 approximately proportional to the quark mass.
 }}
\end{center}
\end{figure}

 The squared ratio of the pion mass to the $\rho$-meson mass is
 plotted in Figure \ref{fig_mPimRho} as a function of $(r_0 m_\pi)^2$,
 both of which are expected to be approximately proportional to the
 quark mass for small quark masses.
 (This holds if the effect of the ``chiral logarithms'' is negligible
 in the quark mass depedence of $m_\pi^2$ and if $r_0$ is approximately
 constant near zero as a function of the quark mass.)
 The straight line in the figure connects the origin and the point
 with the physical values $m_\pi=140{\rm\,MeV}$, $m_\rho=770{\rm\,MeV}$
 and $r_0=0.5{\rm\,fm}$.
 As the figure shows, in this plot there are observable scale breaking
 effects between $\beta=0.67$ and $\beta=0.74$, but the $\beta=0.74$
 points are already close to the continuum expectation.
 Within the (large) statistical errors there is no noticeable difference
 between the points with vanishing and non-vanishing twisted mass.
 (According to Table \ref{tab_chiral_extra} the twisted mass values
 are given by $r_0\mu = 0.02845(68)$ and $r_0\mu = 0.0283(15)$ for
 $\beta=0.67$ and $\beta=0.74$, respectively.)

\section{Twist angle and renormalisation factors}\label{sec3}

\subsection{Twist angle}\label{sec3.1}

 In this section we discuss the determination of the twist angle
 $\omega$.
 For given $(\mu_\kappa,a\mu)$ this is defined as the rotation angle
 relating twisted-mass QCD (TMQCD) to the physical theory QCD.
 An important point is that the connection can be made only after 
 (lattice) renormalisation of the theory.
 The renormalisation of the local bilinears in the Wilson twisted-mass
 formulation is therefore involved.
 Some of the arguments of this section were already discussed 
 in previous publications of this collaboration~\cite{Proc04,DBW}.

 Following~\cite{ALPHA-TMQCD} we operationally define~\cite{Proc04,DBW}
 the twist angle $\omega$ as the chiral rotation angle between the
 renormalised (physical) chiral currents and the corresponding bilinears
 of the twisted formulation.
 We denote with $\hat{V}^a_{x\mu}$ and $\hat{A}^a_{x\mu}$ the 
 physical vector and axialvector currents, while $V^a_{x\mu}$ and
 $A^a_{x\mu}$ are the bilinears of the $\chi$-fields defined in
 Eq.~(\ref{eq05}).
 In order to establish the correspondence with the physical currents,
 the bilinears of the $\chi$-fields have to be properly renormalised.
 This is obtained, as in QCD, by multiplying them by the respective 
 renormalisation constants $Z_V$ and $Z_A$.
 In a mass independent scheme these are functions of $\beta$ alone and
 coincide with the analogous quantities in Wilson lattice QCD for the
 same value of $\beta$.
 So the relation reads:
\begin{eqnarray}\label{eq08}
 \hat{V}^a_{x\mu} &=& Z_V V^a_{x\mu}\,\cos\omega\, + 
\epsilon_{ab} \, Z_A A^b_{x\mu}\,\sin\omega \ ,
\\[0.5em]\label{eq09}
 \hat{A}^a_{x\mu} &=& Z_A A^a_{x\mu}\,\cos\omega\, +
\epsilon_{ab} \, Z_V V^b_{x\mu}\,\sin\omega\,
\end{eqnarray}
 where only charged currents are considered ($a$=1,\,2) and
 $\epsilon_{ab}$ is the antisymmetric unit tensor.

 The {\em conserved} vector current of the $\chi$-fields
\be\label{eq10}
\tilde{V}^a_{x\mu} \equiv 
\frac{1}{4}\left (\overline{\chi}_{x+\mu}\tau_a U_{x\mu}(\gamma_\mu+r)\chi_x +  
\overline{\chi}_{x}\tau_a U^{\dagger}_{x\mu}(\gamma_\mu-r)\chi_{x+\mu}\right )  
\ee
 satisfies by construction the correct Ward-Takahashi identity of the
 continuum.
 In this case the  Formulas~(\ref{eq08}), (\ref{eq09}) apply with
 $Z_V$ replaced by 1, in particular
\be\label{eq11}
 \hat{A}^a_{x\mu} = Z_A A^a_{x\mu}\,\cos\omega\, +
\epsilon_{ab} \, \tilde V^b_{x\mu}\,\sin\omega\ .
\ee

 In practical applications it is useful to define two further angles
 $\omega_V$ and $\omega_A$:
\be\label{eq12}
\omega_V=\arctan (Z_AZ_V^{-1}\tan\omega)\ ,
\hspace{3em}
\omega_A=\arctan (Z_VZ_A^{-1}\tan\omega)\ .  
\ee
 In terms of $\omega_V$, $\omega_A$ Eqs.~(\ref{eq08}) and
 (\ref{eq09}) read
\begin{eqnarray}\label{eq13}
\hat{V}^a_{x\mu} &=& \!\ {\cal N}_V\, (\cos\omega_V V^a_{x\mu} +
\epsilon_{ab}\sin{\omega}_{V} A^b_{x\mu})\ ,
\\\label{eq14}
\hat{A}^a_{x\mu}&=&\!\ {\cal N}_A\, (\cos\omega_A A^a_{x\mu} +
\epsilon_{ab}\sin{\omega}_{A} V^b_{x\mu})\, .
\end{eqnarray}
 The unknown multiplicative renormalisations are now contained in an
 overall factor ($X=V,A$):
\be\label{eq15}
{\cal N}_{X}=\frac{Z_X}
{\cos\omega_X \sqrt{1+\tan{\omega}_V\tan{\omega}_A}}\ .
\ee
 From the definition (\ref{eq12}) it follows 
\begin{eqnarray}\label{eq16}
&&\omega=\arctan \left(\sqrt{\tan{\omega}_V\tan{\omega}_A}\right)\ \\
&&\frac{Z_A}{Z_V}=\sqrt{\tan\omega_V/\tan\omega_A}\label{eq17}\ .
\end{eqnarray}
 As already proposed in~\cite{Proc04,DBW}, we determine the twist angle
 $\omega$ by imposing parity-restoration (up to ${\cal O}(a)$ precision)
 for matrix elements of the physical currents.
 Due to the presence of unknown lattice renormalisations, two conditions
 are required.
 The most suitable choice in the case of the vector current is
\be\label{eq18}
\sum_{\vec{x}} \langle \hat{V}^+_{x0}\, P^-_y\rangle\;=\; 0\ .
\ee
 Indeed, for asymptotic times, the pion state dominates the matrix
 element\footnote{
 At small time-separations, due the ${\cal O}(a)$ breaking of
 parity, intermediate states with ``wrong'' parity may still play a
 role.}
 and the condition reads
\be\label{eq19}
\langle\, 0\,|\, \hat{V}^+_{x0}\,|\,\pi^-\,\rangle\;=\; 0\ .
\ee
 In case of the axialvector current we choose the condition\footnote{
 In~\cite{Proc04,DBW} the use of the {\em temporal} component for the
 currents was proposed.
 This choice is however not optimal: a scalar state with positive parity
 dominates in this case the matrix element in the continuum limit, 
 but at finite lattice spacing the ${\cal O}(a)$ breaking of parity introduces
 contamination by pion intermediate states which eventually dominate for
 light quark masses.}
\be\label{eq20}
\sum_{\vec{x},i} \langle \hat{A}^+_{xi}\, \hat{V}^-_{xi}\rangle\;=\; 0\ 
\ee
 or asymptotically 
\be\label{eq21}
\langle\, 0\,|\, \hat{A}^+_{xi}\,|\,\rho^-\,\rangle\;=\; 0\ .
\ee
 In terms of (\ref{eq13}), (\ref{eq14}) Eqs.~(\ref{eq18}),
 (\ref{eq20}) admit the solution 
\begin{eqnarray}\label{eq22}
\tan{\omega}_V &=&
-i\, \frac{  \sum_{\vec{x}}\langle V^+_{x0}\, P^-_y \rangle}
     {  \sum_{\vec{x}}\langle A^+_{x0}\, P^-_y \rangle} \ ,
\\[1em]\label{eq23}
\tan{\omega}_A &=&
\frac{-i\!\sum_{\vec{x},i}\langle A^+_{xi}\, V^-_{yi}\rangle
         \!+\!\tan{\omega}_V\!
         \sum_{\vec{x},i}\langle A^+_{xi}\, A^-_{yi}\rangle}
     {   \sum_{\vec{x},i}\langle V^+_{xi}\, V^-_{yi}\rangle
         \!+\!i\!\tan{\omega}_V\!
         \sum_{\vec{x},i}\langle V^+_{xi}\, A^-_{yi}\rangle} \ .
\end{eqnarray}
 Eqs. (\ref{eq16}), (\ref{eq17}), (\ref{eq22}) and (\ref{eq23})
 allow the numerical determination of $\omega$ and of the ratio
 $Z_A/Z_V$.

 It is obvious that the definition of the twist angle in the lattice
 theory is subject to ${\cal O}(a)$ ambiguities.
 Different choices of the parity-restoration conditions, including also
 the form of the lattice currents, result in different definitions of
 the twist angle differing by ${\cal O}(a)$ terms.
 The situation of full twist corresponds to
 $\omega=\omega_V=\omega_A=\pi/2$.
 Numerically it is most convenient to use $\omega_V=\pi/2$ as a
 criterion.
 The reason is that a safe determination of the twist angle is obtained
 in the asymptotic regime where the lightest particle dominates as
 intermediate state.
 This is the pseudoscalar state in the case of $\omega_V$ which, as one
 would expect, delivers a better signal than the vector meson in case
 of $\omega_A$.
 Therefore we impose~\cite{Proc04,DBW}
\be\label{eq24}
\omega_V=\frac{\pi}{2} \:\: \Longleftrightarrow\:\:  
\sum_{\vec{x}}\langle A^+_{x0}\, P^-_y \rangle = 0
\ee
 or asymptotically
\be\label{eq25}
\langle\, 0\,|\, A^+_{x0}\,|\,\pi^-\,\rangle\;=\; 0\ 
\ee
 and denote with $\mu_{\kappa cr}$ the corresponding value of
 $\mu_\kappa$ for the given $\mu$.  

 Another possible determination of $\omega_V$ is obtained by replacing
 in~(\ref{eq22}) the currents with their divergences.
 For simplicity, we consider the case of the conserved vector current
 which avoids the introduction of a renormalisation constant:
\be\label{eq26}
\cot{\tilde{\omega}}_V =
i\, \frac
     {  \sum_{\vec{x}}\langle \partial^\ast_\mu A^+_{x\mu}\, P^-_y \rangle}
     {  \sum_{\vec{x}}\langle \partial^\ast_\mu \tilde V^+_{x\mu}\, P^-_y \rangle}
\:=\:\frac{\mcp}{\mu}  \ .
\ee
 Here in the last step~\cite{SHARPE-WU1,SHARPE-WU2} the Ward identity
 for the conserved vector current
 \be\label{eq27}
 \partial^\ast_\mu  \tilde{V}^+_{x\mu} = 2i\mu\, P^+_x
 \ee
 and the definition  (\ref{eq06}) of the ``untwisted'' PCAC quark
 $\mcp$ have been used.
 If the local vector current defined in Eq.~(\ref{eq05}) is used 
 for the determination of $\omega_V$ instead of the conserved one,
 in Eq.~(\ref{eq26}) the introduction of the renormalisation constant
 $Z_V$ is required.
 In this case one has
 \be\label{eq28}
 \cot{\omega}_V =
 i\, \frac
 {  \sum_{\vec{x}}\langle \partial^\ast_\mu A^+_{x\mu}\, P^-_y \rangle}
 {  \sum_{\vec{x}}\langle \partial^\ast_\mu  V^+_{x\mu}\, P^-_y \rangle}
\:=\:Z_V\frac{\mcp}{\mu}  \ ,
 \ee
 where $Z_V$ is determined as explained in the next subsection.
 Using the definition~(\ref{eq12}) for $\omega_V$ one arrives at the
 following relation involving this time the twist angle $\omega$:
 \be\label{eq29}
 \cot{\omega} \:=\:Z_A\frac{\mcp}{\mu}  \ .
 \ee
 Notice that the factor $Z_V$ cancels in this relation which is,
 therefore, independent of the choice for the vector current employed
 for the determination of the twist angle $\omega$.

 One can simply show that the two 
 determinations of $\omega_V$ given by Eqs.~(\ref{eq22}) and 
 (\ref{eq28}) coincide under the assumption that the ratio of the
 correlators is independent of the time separation; this is in 
 particular true for asymptotic times where the pion dominates.

 To have an effective automatic ${\cal O}(a)$ improvement, meaning
 without large ${\cal O}(a^2)$ effects, the critical line
 $(\mu_{\kappa cr}(a,\mu),\mu)$ has to be fixed in such a way that the
 lattice definition of the untwisted quark mass (e.g.~$\mcp$ defined
 above) is free, on that line, from mass independent ${\cal O}(a)$
 errors.
 For a definition of the critical line where this condition is not
 necessarily satisfied, one has to make sure that $\mu > a\Lambda^2$.

 The issue of the choice of the critical untwisted mass has been raised
 by the work of Aoki and B\"ar \cite{AoBa} and by the numerical results
 obtained in \cite{XLF}.
 This problem has been further analyzed in several aspects
 \cite{SHARPE-WU2,FMPR,SHARPE2}.
 In \cite{AoBa,SHARPE-WU2,SHARPE2} the theoretical framework is twisted
 mass chiral perturbation theory (tmChPT) \cite{MUENSTER-SCHMIDT} where
 the cutoff effects are included in the chiral lagrangian along the
 lines of \cite{SHARPE-SINGLETON,RUPAK-SHORESH}.
 The works \cite{AoBa,SHARPE-WU2} agree on the fact that choosing the
 critical mass by imposing $\mcp=0$ (or $\omega_V=\pi/2$) allows to have
 automatic ${\cal O}(a)$ improvement down to quark masses that fulfill
 $\mu \simeq a^2\Lambda^3$.
 In \cite{FMPR} a Symanzik expansion was performed (in an approach
 different from that of refs.~\cite{AoBa,SHARPE-WU2},
 cf.~\cite{SHARPE-WU2} for a discussion) confirming the results of
 \cite{AoBa,SHARPE-WU2}.
 For a discussion of these issues in numerical studies within
 the quenched approximation see~\cite{ZEUTHEN1,REGINA,ZEUTHEN2} and the
 review~\cite{SHINDLER}.

\subsection{Determination of $Z_V$}\label{sec3.2}

 We adopt here the procedure well known in QCD which relies on the
 non-renormalisa\-tion property of the conserved current
 $\tilde V_{x\mu}$~\cite{MaMa}.
 A possible  determination of $Z_V$ in TMQCD is given by
\be\label{eq30}
Z^{(1)}_V=\frac{\langle 0|\tilde V_{x=0,0}^+|\pi^-\rangle}
{\langle 0|V_{x=0,0}^+|\pi^-\rangle}\ .
\ee
 Note that in TMQCD the time component of the vector current couples the
 vacuum to the pseudoscalar particle: in the most interesting region near
 full twist this coupling is maximal.
 (Note that at $a\mu=0$ the analogous procedure has to rely on the
 noisier matrix element with the vector particle or on three point
 functions.)
 Alternatively $Z_V$ can be determined without direct use of the
 conserved current by exploiting the (exact) Ward identity for the
 vector current.
 This implies~\cite{FS}
\be\label{eq31}
\langle 0|\tilde V_{x=0,0}^+|\pi^-\rangle=
\frac{-2i\mu}{m_\pi} \langle 0|P_{x=0}^+|\pi^-\rangle\ .
\ee
 Inserting the above relation in~(\ref{eq30}) a second determination
 of $Z_V$ is obtained:
\be\label{eq32}
Z^{(2)}_V=\frac{-2i\mu\,\langle 0|P_{x=0}^+|\pi^-\rangle}
{m_\pi\,\langle 0|V_{x=0,0}^+|\pi^-\rangle}\ .
\ee
 $Z^{(1)}_V$ and $Z^{(2)}_V$ (differing by ${\cal O}(a)$ terms) are mass
 dependent renormalisations. 
 We obtain a mass independent determination of $Z_V$ by extrapolating
 $Z^{(i)}_V$ to full twist ($\mcp=0$).
 In this situation the theory is ${\cal O}(a)$ improved and the $Z^{(i)}_V$
 deliver an estimate of $Z_V$ with ${\cal O}(a^2)$ error (also including
 ${\cal O}((\mu a)^2)$ terms).

\subsection{Physical quantities}\label{sec3.3}

 The knowledge of the twist angle $\omega$ allows the derivation of
 physical quantities of interest in QCD for a generic choice of
 $(\mu_\kappa,a\mu)$.
 Let us consider the case of the quark mass and the pion decay constant.
 It is convenient~\cite{FS,DMFH,ZEUTHEN1} here to use the conserved vector
 current since it possesses already the right continuum normalisation.
 The {\em physical} PCAC quark mass $\mqp$ can be obtained from the Ward
 identity for the physical axialvector current:
\be\label{eq33}
\langle\partial^*_\mu\hat A^{+}_{x\mu} P^{-}_y\rangle = 
2\amqp \langle P^{+}_{x} P^{-}_y\rangle \ .
\ee
 We use Eq.~(\ref{eq08}) in order to eliminate $A^a_{x\mu}$ in
 (\ref{eq11}) for $\omega \neq 0$
\be\label{eq34}
 \hat A^a_{x\mu}  = -\epsilon_{ab} \hat{V}^b_{x\mu} \cot\omega\, +
\epsilon_{ab} \, \tilde{V}^b_{x\mu} \,(\sin\omega)^{-1}
\ee
 and insert the result in the Ward identity~(\ref{eq33})
 using isospin invariance for $\hat{V}^a_{x\mu}$.
 As a result we obtain:
\be\label{eq35}
\amqp = \frac{-i}{2\sin\omega}
\frac{\langle\partial^*_\mu \tilde V^{+}_{x\mu} P^{-}_y\rangle} 
     {\langle P^{+}_x P^{-}_y\rangle} = \frac{\mu}{\sin\omega}\ ,
\ee
 where in the last step we used once again the Ward
 identity~(\ref{eq27}).
 Inserting Eq.~(\ref{eq29}) into the last expression in the above
 equation, we arrive at the following relation for the untwisted quark
 mass
\be\label{eq36}
\mcp = \mqp Z_A^{-1} \cos \omega \ .
\ee
 In the remainder we shall also make use of a definition of the
 untwisted quark mass which already incorporates the renormalisation
 factor of the axial current:
\be\label{eq37}
\mcpb=\mqp \cos \omega=Z_A \mcp\ .
\ee

 Analogously, for the physical pion decay constant $f_\pi$ we use
\be\label{eq38}
af_\pi=(am_\pi)^{-1} \langle 0|\hat A^+_{x=0,0}|\pi^-\rangle=
-i(am_\pi \sin\omega)^{-1} \langle 0|\tilde V^+_{x=0,0}|\pi^-\rangle\ .
\ee
 Also here the matrix element on the right hand side can be replaced by
 the matrix element of the pseudoscalar density as in~(\ref{eq31})
 giving
\be\label{eq39}
af_\pi = \frac{-2 a\mu}{(am_\pi)^2 \sin\omega} 
\langle 0| P^+_{x=0}|\pi^-\rangle\ .
\ee
 Let us note that here the normalisation of $f_\pi$ corresponds to a
 phenomenological value $\approx 130{\rm\,MeV}$.
 If the local vector current is used in (\ref{eq38}) instead of the 
 conserved one, a factor $Z_V$ is missing:
\be\label{eq40}
af_{v\pi}=-i(am_\pi \sin\omega)^{-1} \langle 0|V^+_{x=0,0}|\pi^-\rangle\ ,\quad
f_{v\pi}=Z_V^{-1}f_{\pi}\ .
\ee
%

\subsection{Results}\label{sec3.4}

\begin{figure}
\begin{center}
\vspace*{0.05\vsize}
\begin{minipage}[c]{1.0\linewidth}
\hspace{0.10\hsize}
\includegraphics[angle=-90,width=0.75\hsize]
 {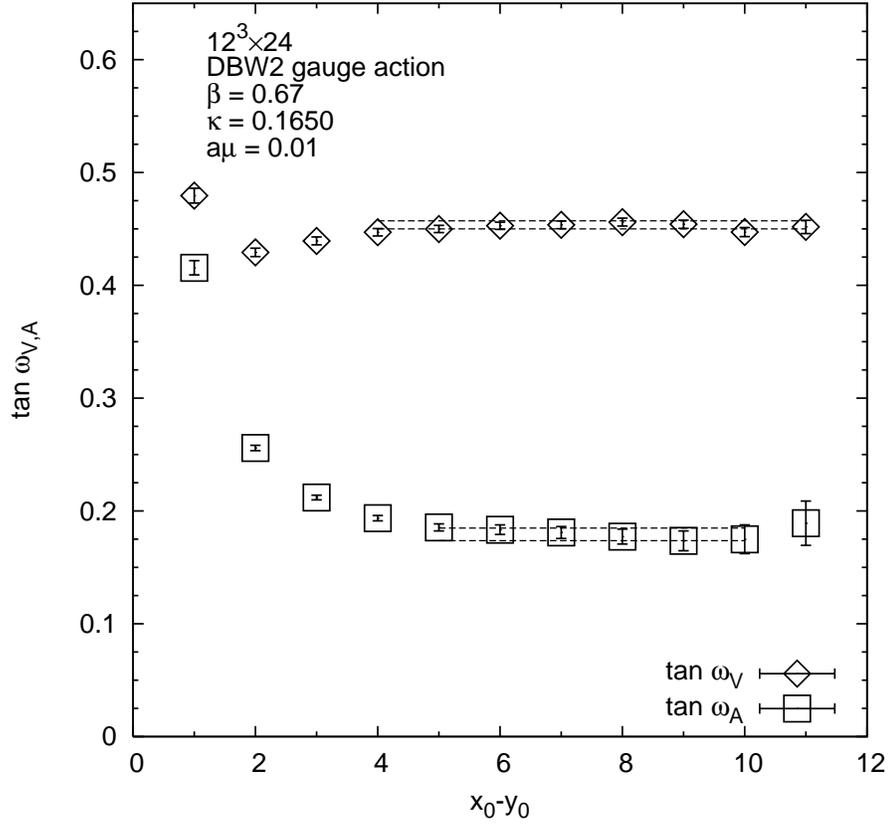}
\end{minipage}
\end{center}
\vspace*{-2em}
\begin{center}
\parbox{0.8\linewidth}{\caption{\label{fig_tanomega}\em
 Determination of $\tan\omega_V$ and $\tan\omega_A$ as in
 Eqs.~(\ref{eq22}), (\ref{eq23})
 for the point ($a^\prime$).
 The lines represent the fitted values.
 }}
\end{center}
\end{figure}

 In Fig.~\ref{fig_tanomega} the local determination of $\omega_V$ and
 $\omega_A$ is shown as a function of the time separation for a specific
 simulation point at positive untwisted quark mass.
 The numerical values of the twist angles $\omega_V$, $\omega_A$ and
 $\omega$ are reported in Table~\ref{tab_omega}.
 Notice that the simulation point at $\beta=0.74$ and $\kappa=0.159$ is
 almost at full twist. 

\begin{figure}
\begin{center}
\vspace*{0.05\vsize}
\begin{minipage}[c]{1.0\linewidth}
\hspace{0.10\hsize}
\includegraphics[angle=-90,width=0.75\hsize]
 {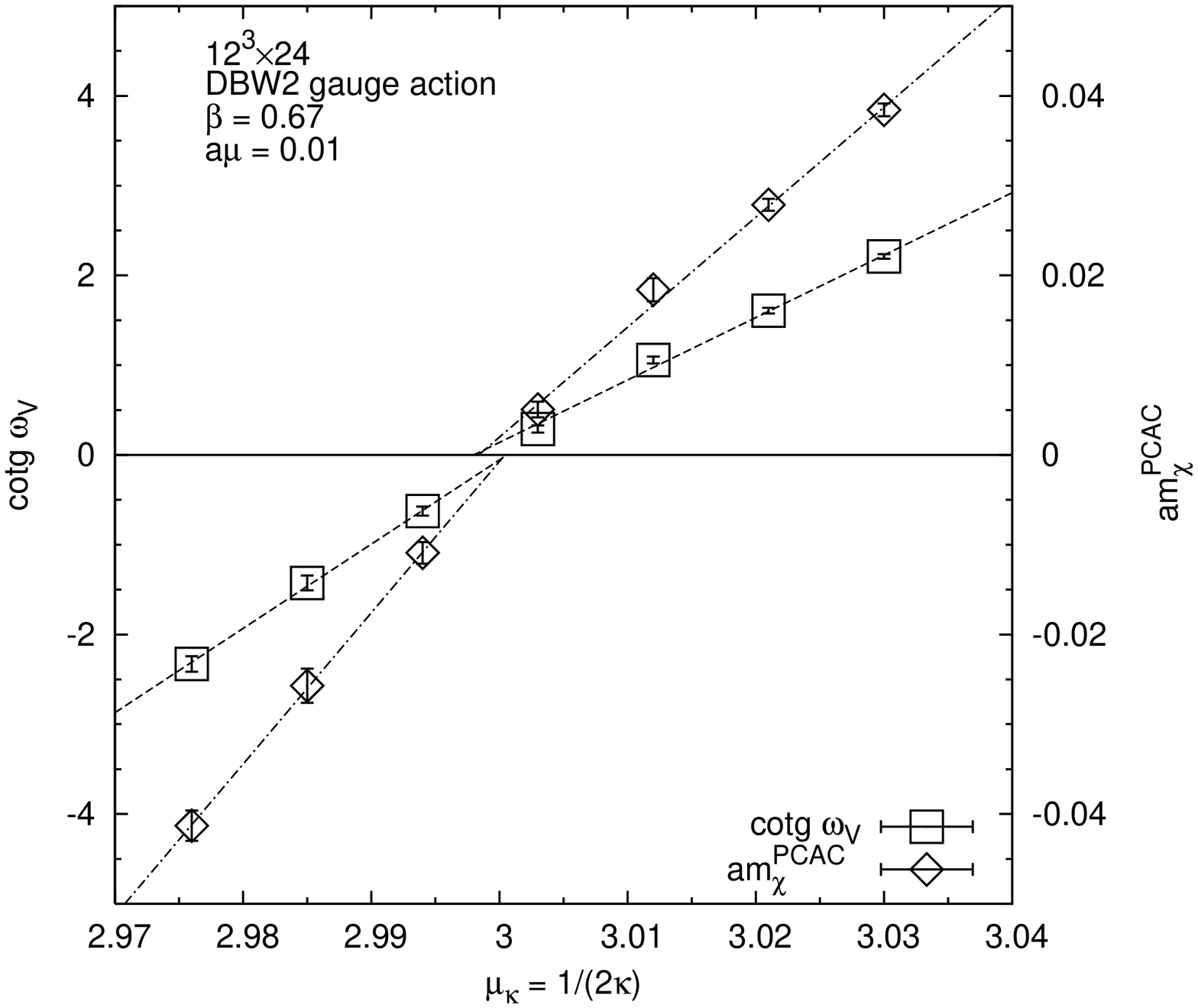}
\end{minipage}
\end{center}
\vspace*{-2em}
\begin{center}
\parbox{0.8\linewidth}{\caption{\label{full_tw_extr_067}\em
 Determination of $\mu_{\kappa cr}$ at $\beta=0.67$, $a\mu=0.01$ by
 parity-restoration and by extrapolating the untwisted PCAC quark mass
 $\mcp$ to zero.
 }}
\end{center}
\end{figure}

\begin{figure}
\begin{center}
\vspace*{0.05\vsize}
\begin{minipage}[c]{1.0\linewidth}
\hspace{0.10\hsize}
\includegraphics[angle=-90,width=0.75\hsize]
 {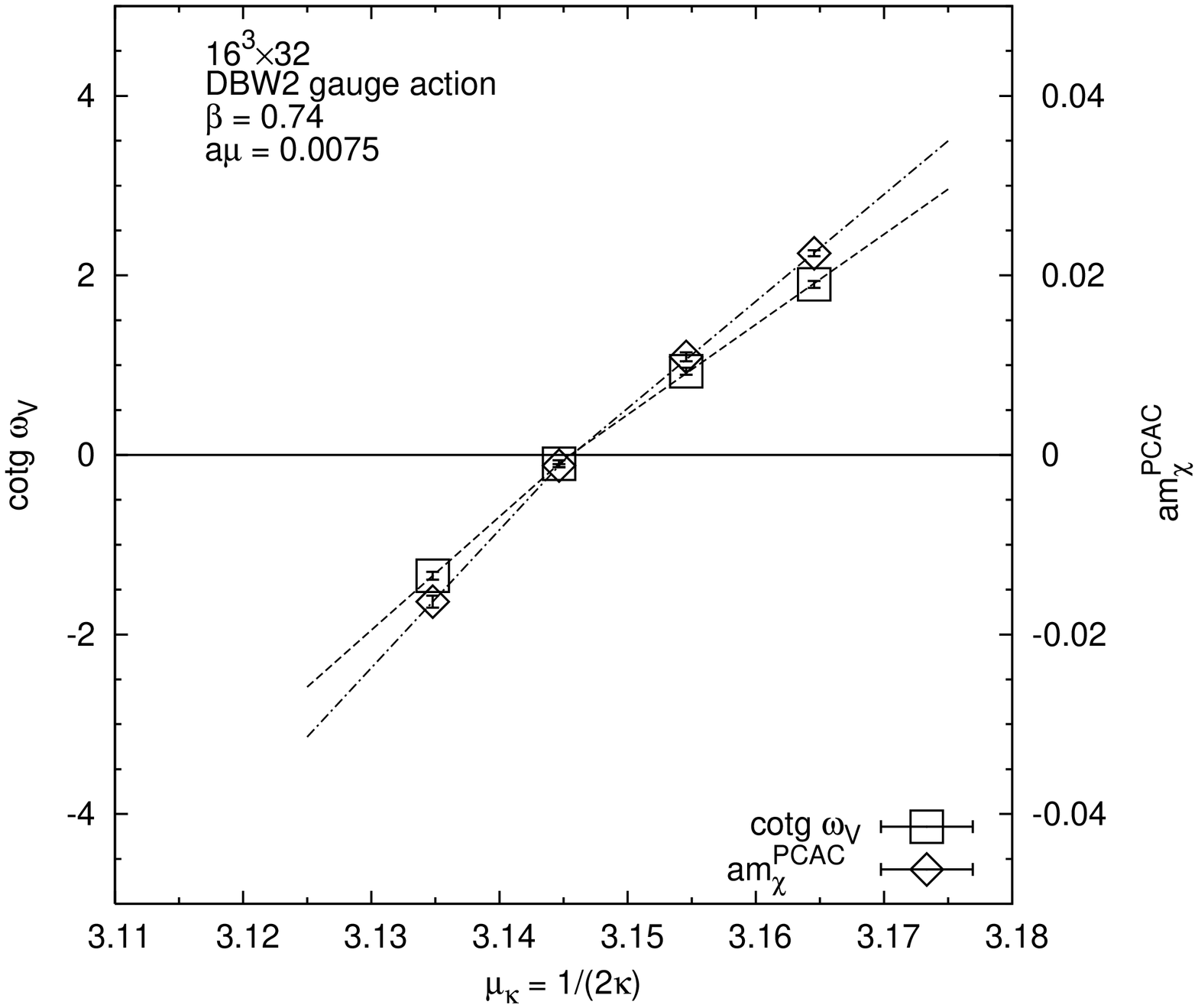}
\end{minipage}
\end{center}
\vspace*{-2em}
\begin{center}
\parbox{0.8\linewidth}{\caption{\label{full_tw_extr_074}\em
 Determination of $\mu_{\kappa cr}$ at $\beta=0.74$, $a\mu=0.0075$ by
 parity-restoration and by extrapolating the untwisted PCAC quark mass
 $\mcp$ to zero.
 }}
\end{center}
\end{figure}

 Figs.~\ref{full_tw_extr_067} and \ref{full_tw_extr_074} show the
 determinations of $\mu_{\kappa cr}$ by extrapolating $\mcp$ and
 $\cot\omega_V$ to zero. 
 The theoretical dependence of the twist angle upon the untwisted 
 bare quark mass $\mu_{\kappa}$ can be obtained~\cite{DBW} by starting
 from the equation~\cite{FrGrSiWe}
\be\label{eq41}
\cot\omega=\frac{m_{\chi R}}{\mu_R} + {\cal O}(a)
\ee
 where $\mu_R$ and $m_{\chi R}$ are the renormalised twisted and
 untwisted quark masses in the continuum limit 
\begin{eqnarray}\label{eq42}
\mu_R &=&  Z_P^{-1} \mu \\
\label{eq43}
m_{\chi R} &=& a^{-1}Z_S^{-1} (\mu_\kappa-\mu_{\kappa cr})\ .
\end{eqnarray}
 Observe that the relation (\ref{eq41}) holds up to ${\cal O}(a)$ terms
 because the right hand side of the relation corresponds to a different
 definition of the twist angle compared to the one given in
 Section~\ref{sec3.1}.
 The two definitions only coincide in the continuum limit. 
 
 By using the first of Eqs.~(\ref{eq12}) one obtains for
 $\omega_V$~\cite{DBW}
\begin{eqnarray}\label{eq44} 
&&\cot\omega_V=(Z_{oV}\mu)^{-1}\, (\mu_\kappa-\mu_{\kappa cr}) + {\cal O}(a)\\
\label{eq45}
&&Z_{oV}=Z_SZ_AZ_P^{-1}Z_V^{-1}\ . 
\end{eqnarray}
 Note that the angular coefficient of the linear fit gives the finite
 combination of renormalisation factors $Z_{oV}$. 
 Using as an input the determination of $Z_A/Z_V$ in Eq.~(\ref{eq17})
 one can obtain from this the combination  $Z_P/Z_S$.

 We use Eq.~(\ref{eq44}) for a linear fit to $\mu_{\kappa cr}$ and
 $Z_{oV}$, see Table~\ref{tab_kcr} for the results.
 As expected from the discussion in Sec.~\ref{sec3.1}, the
 condition $\mcp=0$ gives results very close to those from the
 parity-restoration condition $\cot\omega_V=0$.
 We conclude that the two methods are essentially equivalent also from
 the numerical point of view.
 A discrepancy is observed between the extrapolation from positive and
 negative quark masses for the simulation point $\beta=0.67$: we
 interpret this as a residual effect of the first order phase transition
 at the given value of the lattice spacing.
 (Whether first order phase transition or ``cross-over'' can only be
 decided in a study of the infinite volume limit.)
 Observe also that the $Z_{oV}$ comes out different for the two
 different signs of the quark mass: this is due to the breaking of
 symmetry under reflection of the untwisted quark mass induced by
 ${\cal O}(a)$ terms~\cite{SHARPE2}.
 The numerical discrepancy shows that these ${\cal O}(a)$ corrections are
 relevant.
 An ${\cal O}(a)$-improved estimate of $Z_{oV}$ is simply obtained by averaging
 the determinations for negative and positive quark masses,
 corresponding to a Wilson average for the quantity under study.
%
%
 An analogous observation can be done for other combinations of
 renormalisation constants (see the following).

 Table~\ref{tab_zfac} reports the determination of the renormalisation
 constants of the vector and axialvector currents $Z_V$ and $Z_A$.
 The ratio $Z_A/Z_V$ comes from the analysis of the the twist angles,
 Eq.~(\ref{eq17}).
 Using the direct estimate of $Z_V$ by Eq.~(\ref{eq30}) we can also
 determine $Z_A$. 
 Observe that the full twist extrapolations of $Z_A/Z_V$ from the 
 two quark mass signs present large discrepancies, which in this case
 cannot be attributed to ${\cal O}(a)$ effects (these should disappear at full
 twist). 
 A possible explanation of the discrepancy could reside in the
 relatively bad quality of the data in the negative mass region.
 The discrepancies in $Z_A$ and $Z_P/Z_S$ are a consequence of that
 for $Z_A/Z_V$. In the light of these considerations we rely on the
 determinations for positive quark masses.

\begin{figure}
\begin{center}
\vspace*{0.05\vsize}
\begin{minipage}[c]{1.0\linewidth}
\hspace{0.10\hsize}
\includegraphics[angle=-90,width=0.75\hsize]
 {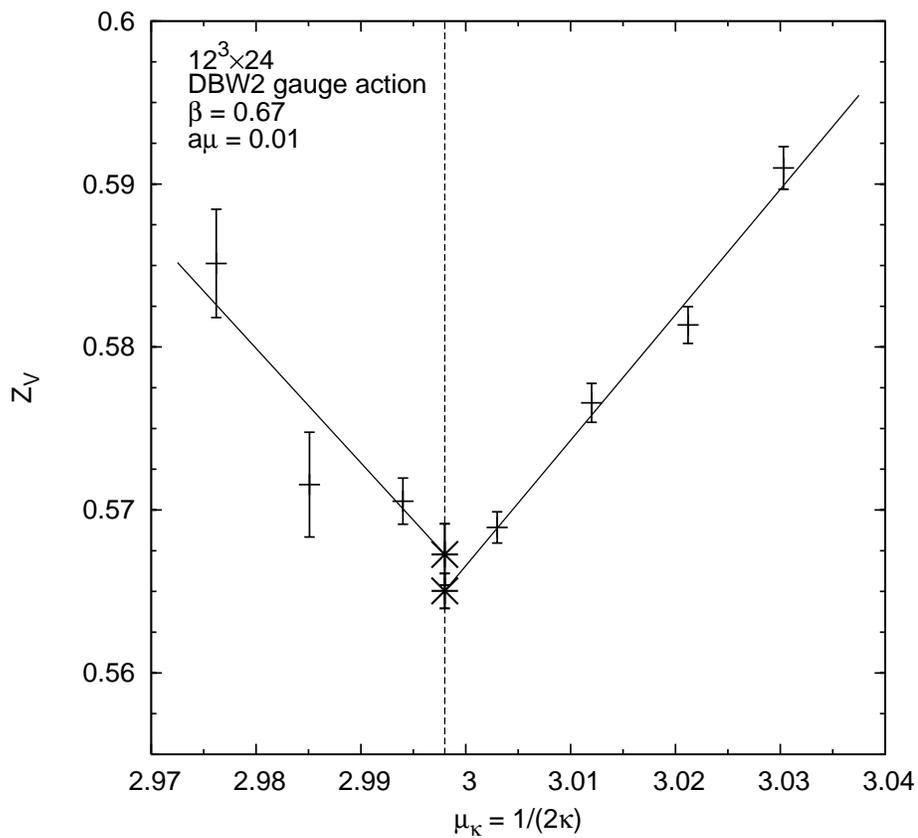}
\end{minipage}
\end{center}
\vspace*{-2em}
\begin{center}
\parbox{0.8\linewidth}{\caption{\label{fig_zv_extr_b067}\em
 Full twist extrapolation of $Z_V^{(1)}$ at $\beta=0.67$, $a\mu=0.01$.
}}
\end{center}
\end{figure}

\begin{figure}
\begin{center}
\vspace*{0.05\vsize}
\begin{minipage}[c]{1.0\linewidth}
\hspace{0.10\hsize}
\includegraphics[angle=-90,width=0.75\hsize]
 {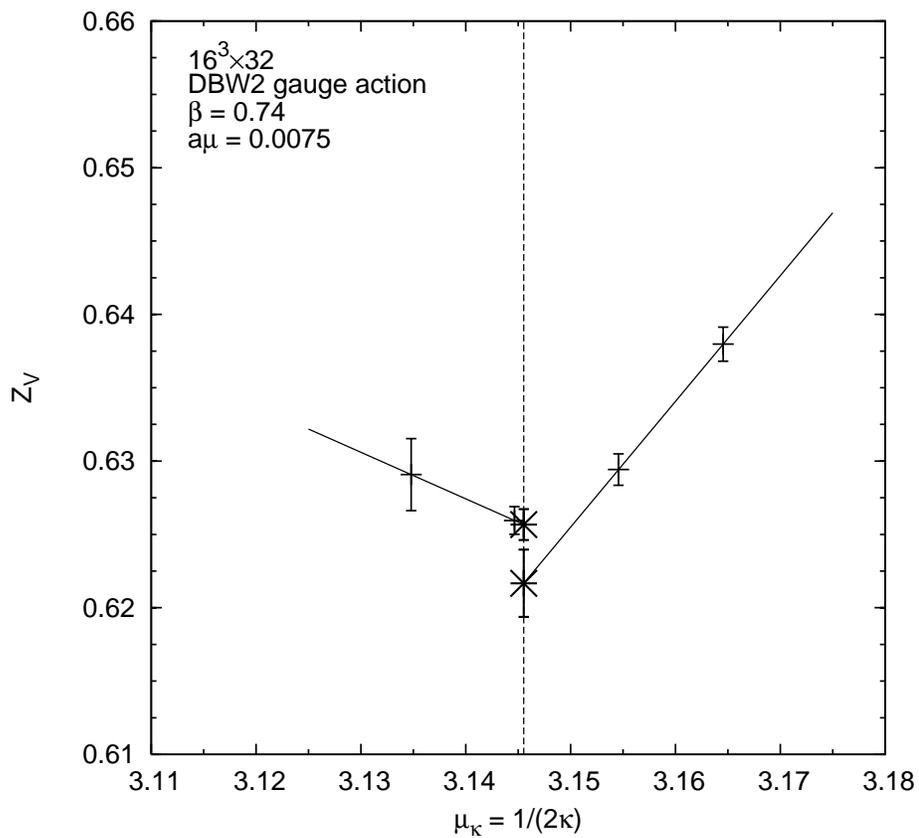}
\end{minipage}
\end{center}
\vspace*{-2em}
\begin{center}
\parbox{0.8\linewidth}{\caption{\label{fig_zv_extr_b074}\em
 Full twist extrapolation of $Z_V^{(1)}$ at $\beta=0.74$, $a\mu=0.0075$. 
}}
\end{center}
\end{figure}

 The full twist extrapolations of $Z_V$ are shown in
 Figs.~\ref{fig_zv_extr_b067} and \ref{fig_zv_extr_b074}: the values
 from the two signs of the quark mass are rather close, compatible with
 each other within statistical uncertainty.
 For the case $\beta=0.74$ the extrapolation is very short, see
 Table~\ref{tab_zfac_ft} for the numerical values with comparison with
 one-loop perturbative estimates~\cite{HP}.
 Table~\ref{tab_zfac_ft} also includes the determinations of the ratio
 $Z_P/Z_S$ from $Z_{oV}$ (see Eqs.~(\ref{eq44}), (\ref{eq45})).
 This quantity is of particular interest for simulations~\cite{progress} 
 of the theory with an additional mass-split doublet
 describing the strange and charm quarks~\cite{FRSD}. 
 Defining $r_{cs}$ as the mass-ratio $m_c/m_s$, the positivity of the
 fermionic measure in the strange-charm sector imposes
\be\label{eq46}
\frac{Z_P}{Z_S}> \frac{r_{cs}-1}{r_{cs}+1}\ .
\ee
 The most stringent condition considering the experimental
 bounds~\cite{PDG04} for  $m_s$ and $m_c$ is
\be\label{eq47}
\frac{Z_P}{Z_S}> 0.89\ .
\ee
 Our results and the tadpole improved perturbative determinations for
 $Z_P/Z_S$ (for $N_f=2$) seem to indicate that already at our values of
 $\beta$ this condition is satisfied.

\begin{figure}
\begin{center}
\vspace*{0.05\vsize}
\begin{minipage}[c]{1.0\linewidth}
\hspace{0.10\hsize}
\includegraphics[angle=-90,width=0.75\hsize]
 {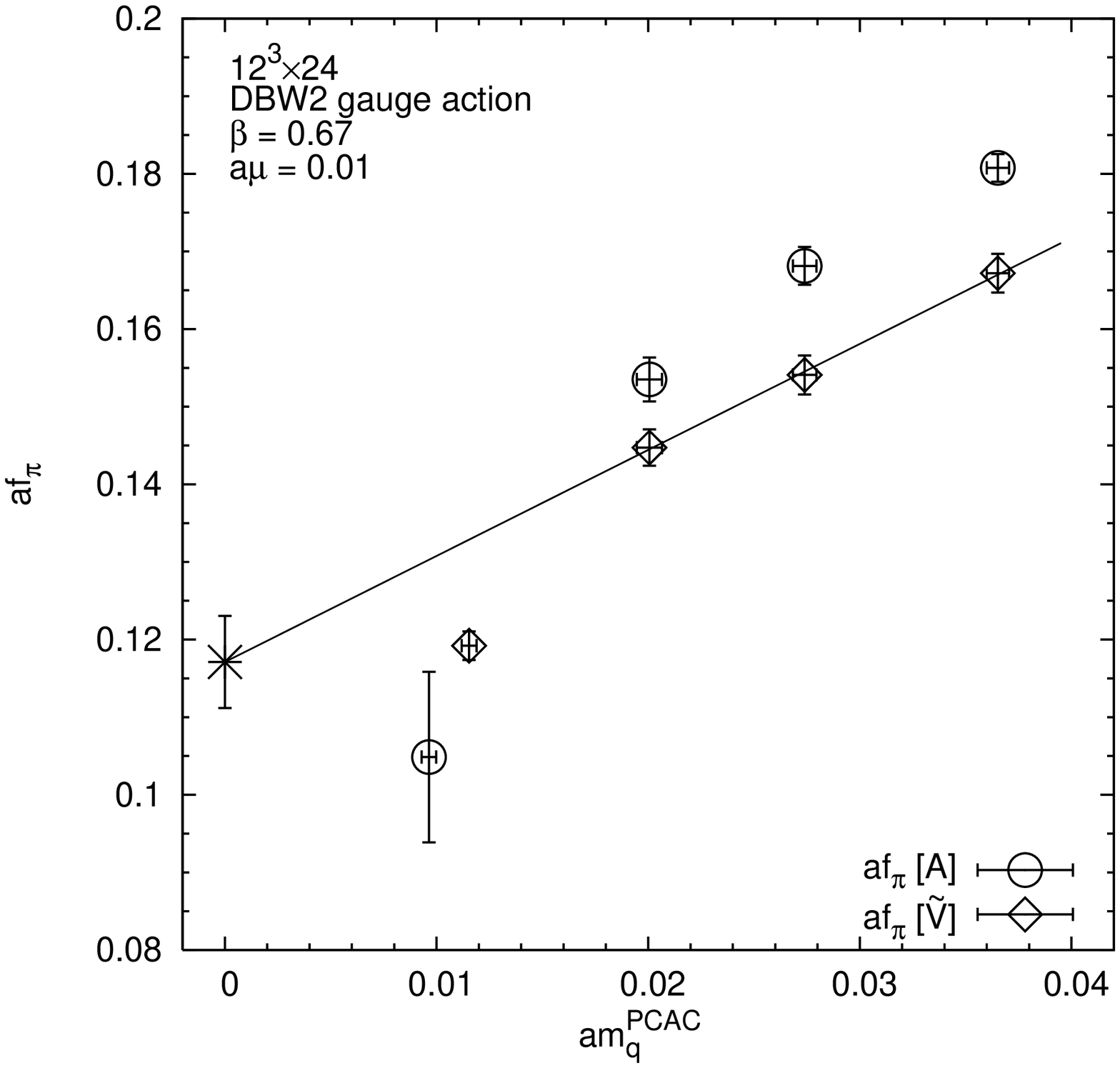}
\end{minipage}
\end{center}
\vspace*{-2em}
\begin{center}
\parbox{0.8\linewidth}{\caption{\label{fig_fp_mq_b067}\em
 The pion decay constant $af_\pi$ as a function of the PCAC quark mass 
 $\amqp$ at $\beta=0.67$, $a\mu=0.01$. 
}}
\end{center}
\end{figure}

\begin{figure}
\begin{center}
\vspace*{0.05\vsize}
\begin{minipage}[c]{1.0\linewidth}
\hspace{0.10\hsize}
\includegraphics[angle=-90,width=0.75\hsize]
 {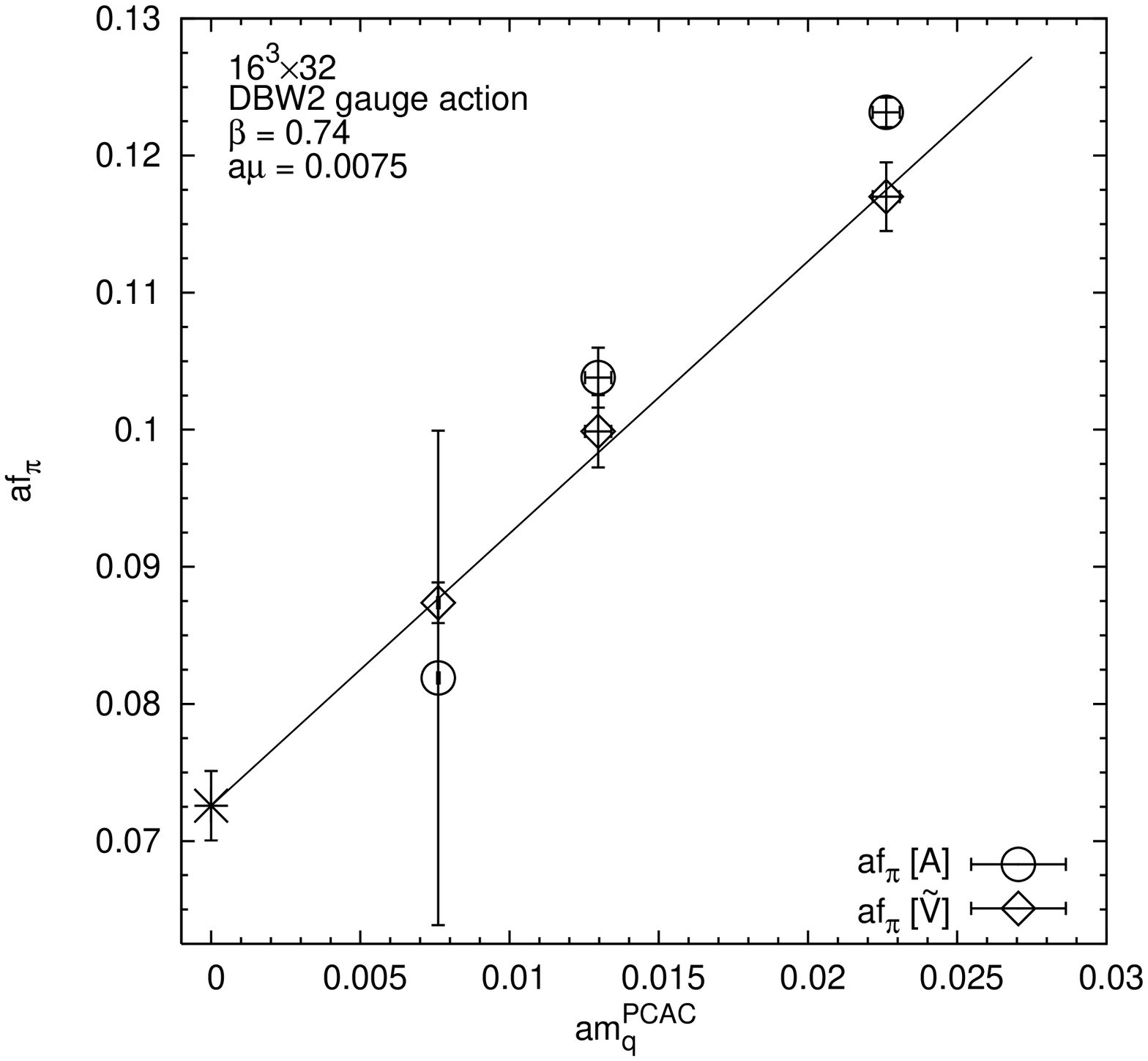}
\end{minipage}
\end{center}
\vspace*{-2em}
\begin{center}
\parbox{0.8\linewidth}{\caption{\label{fig_fp_mq_b074}\em
 The pion decay constant $af_\pi$ as a function of the PCAC quark mass 
 $\amqp$ at $\beta=0.74$, $a\mu=0.0075$. 
}}
\end{center}
\end{figure}

 The results for the physical PCAC quark mass and pion decay constant
 $f_\pi$ obtained from Eqs.~(\ref{eq35}) and (\ref{eq38}) are listed
 in Table~\ref{tab_fpi_mq_phys}.
 In Figs.~\ref{fig_fp_mq_b067} and \ref{fig_fp_mq_b074} the pion decay
 constant is plotted as a function of the quark mass.
 The simulation points for negative quark masses are not taken into
 account in the present discussion.
 The figures also include the determination of $f_\pi$ by the
 axialvector current $A^a_{x\mu}$: a formula similar to
 Eq.~(\ref{eq38}) applies in this case where, however, the factor
 $1/\sin\omega$ is replaced by  $1/\cos\omega$.
 In the interesting region near full twist this introduces large
 fluctuations in the estimate of $f_\pi$, as one can see from the
 figures.
 Moreover in the case of the axialvector current, the decay constant has
 not yet the right normalisation of the continuum: a $Z_A$ factor is
 still missing.
 On the contrary, in the case of the conserved vector current $f_\pi$
 has automatically the physical normalisation~\cite{FS,DMFH,ZEUTHEN1}.
 If we exclude the lightest point at $\beta=0.67$, which is likely to be
 under the influence of residual metastabilities, $f_\pi$ seems to be
 characterised by a linear dependence upon the quark mass.
 On the basis of this observation we try a simple linear extrapolation
 to the chiral limit $\mqp=0$, see Table~\ref{tab_chiral_extra} for the
 numerical results.
 Of course, deviations from this linear behaviour could be present for
 lighter quark masses where chiral logarithms play a role.

 In order to check the scaling between the two $\beta$ values we need to
 fix the lattice spacing.
 This can be accomplished by extrapolating the value of $r_0$ to
 $\mqp=0$.
 Also in this case we obtain two different values for the two different
 signs of the untwisted quark mass, again due to ${\cal O}(a)$ effects.
 As for $Z_{oV}$ we take the average of the two values, which 
 delivers an ${\cal O}(a)$-improved estimate of $r_0$ in the chiral limit. 
 The results are reported in Table ~\ref{tab_chiral_extra}.
 We obtain for the lattice spacing (assuming $r_0=0.5{\rm\,fm}$):
 $a(0.67)=0.1757(41){\rm\,fm}$, $a(0.74)=0.1326(70){\rm\,fm}$.
 Denoting the zero quark mass limit of the pion decay constant by
\be\label{eq48}
f_0 \equiv \lim_{\mqp=0}\; f_\pi \ ,
\ee
 we obtain for $f_0 r_0$: $f_0 r_0(0.67)=0.333(10)$,
 $f_0 r_0(0.74)=0.274(20)$.
 These values are not far from the phenomenological value
 $(f_0 r_0)_{phen}=0.308$.
 (The errors here are only statistical.
 Systematic errors of the chiral extrapolation are not included.)

\section{Fits to chiral perturbation theory}\label{sec4}

 Chiral perturbation theory (ChPT) is an expansion around the limit of
 massless quarks in QCD \cite{CHPT}.
 It describes the dependency of physical quantities on the quark masses
 in terms of expansions in powers of quark masses, modified by
 logarithms.
 In nature, however, quark masses have fixed values.
 The question of how observables depend on them functionally is
 experimentally unaccessible.
 Lattice gauge theory, on the other hand, offers the possibility to vary
 quark masses.
 Therefore it represents the ideal field of application of chiral
 perturbation theory.
 On the one hand, chiral perturbation theory allows to extrapolate
 results from numerical simulations of QCD into the region of small
 physical values for the up- and down-quark masses.
 On the other hand, lattice QCD can provide values for the low-energy
 constants of chiral perturbation theory.

 In chiral perturbation theory the effects of the non-zero lattice
 spacing $a$ can be taken into account in form of an expansion in powers
 of $a$\cite{SHARPE-SINGLETON,LEE-SHARPE,RUPAK-SHORESH,AOKI-OA,BRS}.
 For the case of the Wilson twisted-mass formulation of lattice QCD this
 has been worked out in next-to-leading order in
 \cite{MUENSTER-SCHMIDT,MSS,SCORZATO,SHARPE-WU2}.

 The major purpose of the present paragraph is to provide a set of 
 formulas derived from lattice chiral perturbation theory that can be
 used to analyze physical quantities such as the pion mass, decay
 constants and amplitudes.
 The novelty here is that these quantities have to be described 
 {\em across} or nearby a phase transition.

 The ChPT formulas are expected to be applicable at sufficiently small
 values of the lattice spacing and quark mass.
 It is thus far from obvious whether the data obtained with the DBW2
 action in this work can be described by them, hence it is interesting
 to confront the simulation data at our quark masses and lattice
 spacings with these formulas.
 Let us emphasize that we consider this investigation mainly as a
 methodological study that does not aim to extract physical values of
 the low energy constants in the first place.

 Properly determined parameters of the ChPT formulas in the continuum
 limit are independent of the lattice action.
 The parameters describing the dependence on the lattice spacing do,
 however, depend on it.
 Therefore, in an Appendix we also present ChPT fits of some simulation
 data obtained previously with the Wilson plaquette gauge action
 \cite{PLAQUETTE}. 

 The quark masses in chiral perturbation theory always appear multiplied
 by $2 B_0$, where $B_0$ is a low-energy constant.
 A connection to lattice regularisation can be established by
 considering the renormalised quark masses defined in
 Eqs.~(\ref{eq42}), (\ref{eq43}) and
\begin{equation}\label{eq49}
\mcpr = \frac{Z_A}{Z_P} \mcp\, .
\end{equation}
 A common renormalisation factor $1/Z_P$ in $\mcpr$ and $\mu_R$ can be
 absorbed into $B_0$.
 However, since the multiplicative renormalisation of $\mcp$ and $\mu$
 differs by a factor $Z_A$, this has to be taken into account when
 fitting lattice data (see below).

 The lattice spacing enters chiral perturbation theory in the
 combination
\begin{equation}\label{eq50}
\rho = 2 W_0 a\,,
\end{equation}
 where $W_0$ is another low-energy constant.

 For the low-energy constants of lattice QCD in next to leading order
 \cite{RUPAK-SHORESH,BRS} with two quark flavours we use the notation
\begin{equation}\label{eq51}
L_{54} = 2 L_4 + L_5\,,\quad
L_{86} = 2 L_6 + L_8\,,\quad
W_{54} = 2 W_4 + W_5\,,\quad
W_{86} = 2 W_6 + W_8\,,
\end{equation}
\begin{equation}\label{eq52}
W = \half (W_{86} - 2 L_{86})\,,\quad
W^\prime = \half(W^\prime_{86} - W_{86} + L_{86})\,,\quad
\widetilde{W} = \half (W_{54} - L_{54})\,.
\end{equation}

 Experience in untwisted lattice QCD shows \cite{QQ+Q} that lattice
 artefacts are considerably reduced when observables are considered as
 functions of the PCAC quark mass instead of the renormalised lattice
 quark mass.
 (A possible reason is that the PCAC quark mass reabsorbs leading
 order {\cal O}(a) effects.)
 Therefore, in our case, instead of using $m_{\chi R}$ as a variable, we
 re-expand the physical quantities in terms of the PCAC quark mass in
 the twisted basis $\mcpr$.
 Including the relevant prefactor we define
\begin{equation}\label{eq53}
\cpp = 2 B_0 \mcpr\,.
\end{equation}
 For the purpose of fitting data at constant $\mu$ it is convenient to
 define the combination
\begin{equation}\label{eq54}
\bar{\chi} = 2 B_0 \sqrt{(\mcpr)^2 + \mu_R^2}\,.
\end{equation}
 (The attentive reader is certainly realising that we use the symbols
 $\chi$ for different quantities.
 Nevertheless, both the notation for the fermion field of twisted-mass
 fermions and the mass parameters in ChPT are standard in the
 literature and we do not want to change neither of them in this paper.)
 Then, for the charged pion masses, chiral perturbation theory at
 next-to-leading order including lattice terms of order $a$ gives
\begin{equation}\label{eq55}
m_{\pi\pm}^2 = \bar{\chi}
+ \frac{1}{32 \pi^2 F_0^2} \bar{\chi}^2
\ln \frac{\bar{\chi}}{\Lambda^2}
+ \frac{8}{F_0^2}
\{ (- L_{54} + 2 L_{86}) \bar{\chi}^2
+ 2 (W - \widetilde{W}) \rho\,
\cpp \}\,.
\end{equation}
 Similarly for the pion decay constant and the one-pion matrix element
 of the pseudoscalar density:
\begin{equation}\label{eq56}
\frac{F_\pi}{F_0} = 1
- \frac{1}{16 \pi^2 F_0^2} \bar{\chi} \ln \frac{\bar{\chi}}{\Lambda^2}
+ \frac{4}{F_0^2} \{ L_{54} \bar{\chi}
+ 2 \widetilde{W} \rho\,
\frac{\cpp}{\bar{\chi}} \}\,,
\end{equation}
\begin{equation}\label{eq57}
\frac{G_\pi}{F_0 B_0} = 1
- \frac{1}{32 \pi^2 F_0^2}
\bar{\chi} \ln \frac{\bar{\chi}}{\Lambda^2}
+ \frac{4}{F_0^2} \{ (- L_{54} + 4 L_{86}) \bar{\chi}
+ ( 4 W - 2 \widetilde{W}) \rho\,
\frac{\cpp}{\bar{\chi}} \}\,.
\end{equation}
 In the ChPT formulas the pion decay constant at zero quark mass ($F_0$)
 appears.
 In the conventional normalisation its phenomenological value is
 $F_0 \approx 86{\rm\,MeV}$.
 This is related to $f_0 \approx 122{\rm\,MeV}$ used in the previous
 section by $F_0 \equiv f_0/\sqrt{2}$.
 Similarly, $F_\pi$ and $G_\pi$ denote the pion decay constant and the
 one-pion matrix element of the pseudoscalar density, respectively, in
 this normalisation convention.

 The renormalisation scale $\Lambda$ appearing in the one-loop
 contributions is taken to be $\Lambda = 4\pi F_0$ as usual.
 Taking into account the renormalisation factors, when using these
 expressions for fitting the lattice data, one writes
\begin{equation}\label{eq58}
\bar{\chi}
= 2 B \sqrt{(\mcp)^2 + Z_{A}^{-2} \mu^2}\,,
\end{equation}
 where $B = B_0 Z_A / Z_P$.

\subsection{Fit procedure}\label{sec4.1}

 For fitting the data as a function of $\mcp$ Eqs.~(\ref{eq55}),
 (\ref{eq56}) and (\ref{eq57}) are going to be used.
 The data for $m_{\pi}$, $F_\pi$ and $G_\pi$, as well as that for
 $\mcp$ are afflicted with numerical errors.
 Therefore, a fit procedure has to be used which takes into account
 errors in both coordinates.
 The method with effective variances \cite{OREAR} treats the coordinates
 on unequal footings but is numerically not so convenient.
 We have decided to use the more appropriate method of \emph{generalised
 least-squares fits} \cite{MARSHALL-BLENCOE}.

 Consider a data set containing $N$ ``measured'' values for each of
 the $D$ variables.
 They are collected in the vector
 $\mathbf{y}=\left(\mathbf{y_{1}},\dots\mathbf{,y_{N}}\right)$, where
 each element $\mathbf{y_{i}}$ is itself a column vector with $D$
 elements $\mathbf{y_{i}}=\left\{ y_{i,j}\right\},\ j=1,\dots,D$.
 The true values for each data point, which have to be estimated
 together with the parameters, will be collected in the same way in a
 vector $\mathbf{x}=\left(\mathbf{x_{1},\dots,x_{N}}\right)$ with
 entries $\mathbf{x_{i}}=\left\{ x_{i,j}\right\},\  j=1,\dots,D$.
 Now the set of measured data points $\left\{ y_{i,j}\right\} $
 represents a single realization of an experiment which occurs with a
 probability given by a joint distribution called ``likelihood''.
  The likelihood is specified by a multivariate normal distribution $L$
 with mean values given by the exact values $\mathbf{x}$ and a
 $ND\times ND$ covariance matrix
 $\sigma=\left\{ \sigma_{(i,j),(k,l)}\right\},\ i,k=1,\dots,N;\
 j,l=1,\dots,D$:
\begin{equation}\label{eq59}
L=\frac{1}{\left(2\pi\right)^{\frac{ND}{2}}}\frac{1}{\sqrt{\det\sigma}}
\exp\left[-\frac{1}{2}\left(\mathbf{x}-\mathbf{y}\right)\sigma^{-1}
\left(\mathbf{x}-\mathbf{y}\right)^{T}\right].
\end{equation}

 The process of data analysis amounts to the constrained maximisation of
 this likelihood through the estimation of the values of $\mathbf{x}$
 based on the knowledge of $\mathbf{y}$, where the constraints enter
 through the fit-functions.
 Instead of maximising $L$ it is more convenient to minimise its
 negative logarithm.
 The only non-constant term is given by
\begin{equation}\label{eq60}
L'=\frac{1}{2}\left(\mathbf{x}-\mathbf{y}\right)\sigma^{-1}
\left(\mathbf{x}-\mathbf{y}\right)^{T}.
\end{equation}

 The fit-functions are given by a number $F$ of model-functions $G_{i}$,
 which can be incorporated as, generally nonlinear, constraints on the
 relationship between the exact values collected in $\mathbf{x}$.
 These functions also depend on a set of $P$ parameters
 $\alpha=\left(\alpha_{1},\dots,\alpha_{P}\right)$, whose values are to
 be determined.
 They can be written in the compact form
 $\mathbf{G}\left(\mathbf{x},\alpha\right)=\mathbf{0}$ with the
 $F$-dimensional column vector
 $\mathbf{G}=\left(G_{1},\dots,G_{F}\right)$.

 Maximisation of the likelihood $L$ under the constraints
 $\mathbf{G}\left(\mathbf{x},\alpha\right)=\mathbf{0}$ is now equivalent
 to the unconstrained minimisation of $\mathcal{L}$ given by
\begin{equation}\label{eq61}
\mathcal{L}=\frac{1}{2}\left(\mathbf{x}-\mathbf{y}\right)\sigma^{-1}
\left(\mathbf{x}-\mathbf{y}\right)^{T}+\lambda\mathbf{G}\,,
\end{equation}
 where $\lambda$ is the $F$-dimensional row-vector of Lagrange multipliers.
 We implemented the minimisation of $\mathcal{L}$ using the Maple
 algorithm NLPSolve, which is based on routines provided by the Numerical
 Algorithms Group (NAG).

 In the present case the $N$ different points of measurement correspond
 to different values of the hopping parameter $\kappa$, which are
 completely independent of each other.
 Therefore we can assume the covariance matrix to be diagonal,
 $\sigma_{(i,j),(i,j)}=\left(\Delta y_{i,j}\right)^{2}$, where
 $\Delta y_{i,j}$ denotes the statistical error of $y_{i,j}$.

 The errors of the model-parameters $\alpha_{i}$ are calculated using a
 Monte-Carlo approach.
 In $K$ steps of an artificial Monte-Carlo procedure a new set of
 normally distributed values
 $\left\{y_{i,j}^{mc}\right\}_{k},\ k=1,\dots,K$, is generated using the
 values of $\left\{ y_{i,j}\right\}$ as means and $\sigma_{i,j}$ as the
 variances.
 Now for every $k$ an independent estimate for the parameters is
 calculated yielding $\alpha_{mc;i}^{k}$ in each step.
 Finally the errors $\Delta\alpha_{i}$ are given by the standard
 deviation of the set of
 $\left\{ \alpha_{mc,i}^{k}\right\},\ k=1,\dots,K$.

\subsection{Results}\label{sec4.2}

 At $\beta = 0.67$ and at $\beta = 0.74$ results for $m_{\pi}$,
 $F_\pi$, $G_\pi$ and $\mcp$ are available both for non-vanishing and
 for vanishing twist mass $\mu$.
 At $\mu = 0$ only part of the data, namely for $m_0 - m_{cr} > 0$, is
 reliable and is being used.

 By using the results in Table \ref{tab_chiral_extra} for the values of
 $r_0/a$ extrapolated to the chiral limit, we express all quantities
 in units of MeV.
 For the value of the Sommer scale we assume
 $r_0 \equiv 0.5\ \textrm{fm} = (394.6..\ \textrm{MeV})^{-1}$.
 This allows us to compare and to combine the results from the different
 values of $\beta$.

 It is important to observe that the lattice spacing $a(\beta)$ is
 obtained from extrapolation of $r_0/a$ to the chiral limit.
 In presence of both positive and negative masses we take the average.
 This is a strong constraint on the fits, since the data have to
 reproduce the scaling behaviour dictated by $r_0$.
 If the purpose is the determination of the low-energy constants,
 matching ratios like $m_{\pi}/F_\pi$ with ChPT would be preferable.
 However, in this  exploratory study, we find it interesting to check
 that the scaling behaviour of different quantities is indeed
 consistent.

 We made combined fits of the three quantities as functions of $\mcp$
 for both values of $\beta$, including lattice terms of order $a$.
 For the pion masses the expressions for the $\mathcal{O}(a^2)$ lattice
 terms are known, but they cannot be fitted meaningfully.
 The value of $Z_A$, entering the fit functions, has been taken as input
 from the Monte Carlo data.
 As it varies with $\beta$, we denote the corresponding values
 $Z_A(\beta)$.
 The fits include data both for non-zero and zero twisted mass $\mu$.

 The low-energy constants resulting from the fits are shown in Table
 \ref{fit_resu_dbw}.
 In the first case data points with both positive and negative values
 of $\mcp$ are fitted, whereas in the second case only those with
 $\mcp>0$.
 (This latter choice corresponds to the procedure in Section~\ref{sec3}
 where also only points with $\mcp>0$ have been taken into account in
 the chiral extrapolation of $f_{\pi}$.)
 We also made single fits for the three quantities, which are not
 displayed here.
 As they are each based on less data, their results are less valuable,
 but consistent with the combined fits.

 In addition to the single-$\beta$ fits we also made a global fit
 including the data from both values of $\beta$.
 The results are also contained in Table \ref{fit_resu_dbw}.
 The fits at the two single values of $\beta$ and the global fit are
 roughly consistent with each other.
 The differences in the numbers for the low-energy constants give an
 indication of the size of the uncertainties.

 Instead of using $Z_A$ as input from the numerical calculations, it can
 alternatively be left as an additional fit parameter.
 The corresponding fit results are shown in the right hand side of the
 table.
 The fitted $Z_A$ is in rough agreement with its Monte Carlo estimate.
 Also, the low-energy coefficients are consistent with the ones from the
 other fits.

 In addition to the combinations of Gasser-Leutwyler coefficients $L_k$,
 the table includes the values of the invariant scale 
 parameters~\cite{LEUTWYLER}
\begin{eqnarray}\label{eq62}
\Lambda_3 &=& 4\pi F_0 \exp (128 \pi^2 (L_{54} - 2 L_{86}))\,, 
\nonumber
\\[0.5em]
\Lambda_4 &=& 4\pi F_0 \exp (32 \pi^2 L_{54})\,.
\end{eqnarray}

 The results for $\Lambda_3$, $\Lambda_4$ are close to phenomenological
 estimates (see the discussion).
 The $W$ parameters have large errors but their magnitude is reasonable,
 as $W_0$ is expected to be of order $\Lambda_{QCD}^3$ and the other
 $W$'s of the same order as the $L$'s.

 The fit curves for $m_{\pi}$, $F_\pi$ and $G_\pi$ together with
 data points at $\beta=0.67$ and $\beta=0.74$ are shown in
 Figs.~\ref{mpi-fit-067}, \ref{mpi-fit-074}, \ref{fpi-fit} and
 \ref{gpi-fit}.
 In order to display the size of the leading order contribution and the
 corrections, the figures contain additional curves representing the fit
 functions with some of the low-energy constants being put to zero.

 We have also investigated $\mcp$ as a function of $m_0$.
 It can be fitted with the corresponding formula from chiral
 perturbation, which involves $W$, $W^\prime$ and $\widetilde{W}$ but no
 $L$-coefficients, but the resulting coefficients are unreliable owing
 to large errors.

 In this section we stick to the definition of the untwisted bare
 PCAC quark mass $\amcp$ in Eq.~(\ref{eq06}). 
 As it is shown in Appendix \ref{appendix}, the agreement with the
 ChPT formulas can be improved by taking $\amcpb$ of Eq.~(\ref{eq37})
 as the quark mass variable, instead.
 In addition, ChPT fits to some previously obtained simulation data
 by the Wilson plaquette gauge action are also presented there.

\section{Discussion}\label{sec5}

 We compared in this paper the numerical simulation results with two
 flavours of twisted-mass Wilson quarks and DBW2 gauge action at two
 values of the lattice spacing corresponding to $\beta=0.67$ and
 $\beta=0.74$.
 The lattices were $12^3 \cdot 24$ and $16^3 \cdot 32$, respectively.
 The lattice spacing was defined by the value of the Sommer scale
 parameter $r_0$ extrapolated to zero quark mass and assuming
 $r_0 \equiv 0.5{\rm\,fm}$.
 The $\beta$-values were chosen in such a way that the lattice
 extensions were approximately equal: $L \simeq 2.11{\rm\,fm}$ and
 $L \simeq 2.12{\rm\,fm}$, respectively.
 Also the bare twisted masses scaled approximately:
 $r_0\mu\simeq 0.0285$ and $r_0\mu\simeq 0.0283$, respectively.

 The comparison of the observed quantities at two $\beta$-values
 allows for a first look at discretisation errors.
 The outcome of these tests is reasonable, having in mind the coarse
 lattice spacings: $a \simeq 0.176{\rm\,fm}$ on the
 $12^3 \cdot 24$ and $a \simeq 0.133{\rm\,fm}$ on $16^3 \cdot 32$.
 For instance, the results for the pseudoscalar decay constant at zero
 quark mass are $f_\pi r_0 = 0.330(10)$ and $f_\pi r_0 = 0.274(20)$
 at $\beta=0.67$ and $\beta=0.74$, respectively.
 These values also come close to the phenomenological value
 $(f_\pi r_0)_{phen}=0.308$ \cite{DURR}.
 The situation is somewhat worse for the pseudoscalar-vector mass ratio,
 as Figure~\ref{fig_mPimRho} indicates.
 There are some noticeable scale breaking effects, especially for
 pseudoscalar masses near $m_\pi=r_0^{-1}$.
 Of course, one has to have in mind that the $\rho$-meson mass in most
 of the points is quite close to the cutoff.

 The prerequisite for the extraction of quantities as, for instance,
 $f_\pi$ is the know\-ledge of the multiplicative renormalisation
 $Z$-factors for the currents.
 For obtaining the $Z$-factors one can exploit the twist-angle
 dependence in the plane of untwisted and twisted quark masses.
 As we have shown in Section \ref{sec3}, this is a rather powerful
 method for obtaining ``finite'' (according to perturbation theory)
 $Z$-factor combinations as $Z_V$, $Z_A$ and $Z_P/Z_S$.
 Remarkably consistent results could be obtained even with our
 exploratory simulation data, without a dedicated choice of simulation
 points for this purpose.

 We have also attempted to describe our numerical simulation data by a
 set of formulas derived from lattice chiral perturbation theory.
 Although the values of the lattice spacing and the quark mass are
 rather large in the simulations, it turned out that these formulas
 describe the behaviour of many physical quantities -- even across the
 phase transition -- surprisingly well, at least on a qualitative level.
 However, at the quantitative level our presently available data do
 neither allow to make a quantitative extraction of the values of the
 ChPT parameters nor can we answer the question whether the lattice
 artifacts are well described by the lattice extension of ChPT.
 The achieved qualitatively correct ChPT fits of our simulation
 data makes us very optimistic that with new data we are working on
 at present -- at smaller lattice spacings and small quarks masses --
 these questions will be answered.
 To achieve this the experience with the fits in this paper will be
 very helpful.

 In Section \ref{sec4} and Appendix \ref{appendix} we used the NLO
 expressions of ChPT including terms describing ${\cal O}(a)$
 lattice artefacts.
 In general, metastable points near the first order phase transition
 can be and have been included in the fits.
 (Note that the fits in Ref.~\cite{QQ+Q} are also based on metastable
 points, as it has been discovered later.)
 Several setups were tried and were shown to give satisfactory and
 consistent fits.
 Nevertheless, there are probably some higher order effects (higher
 orders both in the quark mass and in lattice spacing) which are
 non-negligible in our parameter range.
 In addition, for the multi-parameter fits our data are not precise
 enough and the data points are too few and not optimally distributed
 in the parameter space.
 (In a dedicated investigation the inclusion of partially quenched
 data points could be very helpful.)
 Qualitatively speaking, the ChPT fits presented here support
 the choice of the PCAC quark mass as the preferred quark mass
 variable and show that the ${\cal O}(a)$ effects are not overwhelming
 because a fit without them is most of the time possible.
 Both these findings agree with those of Ref.~\cite{QQ+Q}.

 The ChPT fits are also helpful in estimating the minimal pion mass
 at a given lattice spacing.
 For instance, the results at $\beta=0.74\;(a=0.1326(70){\rm\, fm})$
 indicate that for fixed $a\mu=0.0075$ we are above the end point
 of the first order phase transition line (see e.g. the smooth
 behaviour near $\mu_{\kappa cr}$ in Figure~\ref{full_tw_extr_074}).
 The minimal value of the pion mass in Figures~\ref{mpi-fit-074} and
 \ref{fig_mpi_dbw} is about
 $m_\pi^{min}(a\mu=0.0075) \simeq 280{\rm\,MeV}$.
 This is an upper bound for the absolute minimum $m_\pi^{min}$ at
 $\beta=0.74$.

 According to Table~\ref{fit_resu_dbw} the fits of the data with DBW2
 gauge action suggest the following qualitative estimates for the
 values of the relevant ChPT parameters:
\begin{eqnarray}\label{eq63}
& 2.9\,{\rm GeV} \leq B \leq 3.5\,{\rm GeV}  \nonumber\\
& 70\,{\rm MeV} \leq F_0 \leq 85\,{\rm MeV}  \nonumber\\
& 4.0 \leq \Lambda_3/F_0 \leq 8.0            \nonumber\\
& 16.0 \leq \Lambda_4/F_0 \leq 19.0
\end{eqnarray}
 As Table~\ref{fit_resu_plq} shows, the fits with the plaquette gauge
 action are roughly consistent with these values.
 The estimates of $\Lambda_{3,4}$ are close to previous estimates
 in \cite{QQ+Q}: $\Lambda_3/F_0 \approx 8$, $\Lambda_4/F_0 \approx 21$.

 The values of the $W$-parameters describing ${\cal O}(a)$ effects
 are not well determined and are in most cases consistent with zero in
 our fits, if $\amcp$ (or $\amcpb$) is taken as the independent
 variable.
 Note that if one considers the relation of $\amcp$ and $am_0$ then 
 $W$ and $W^\prime$ are quite visible.
 An example is Figure 2 in our previous proceedings contribution
 \cite{CYPRUS} where $W$ gives the difference of the slope between
 positive and negative masses ($W^\prime$ comes out to be small).

 If the data at the two $\beta$-values are fitted separately, as
 Table~\ref{fit_resu_dbw} shows, there is a remarkably good agreement
 of the corresponding parameter values.
 This agrees with expectations since the inclusion of ${\cal O}(a)$
 terms in the formulas reduces the discretisation errors in the physical
 parameters.
 The consistency of the ChPT fits is supported by the agreement of the
 pion decay constant at zero quark mass $F_0$ with the value directly
 extracted from the data in Section~\ref{sec3}: 
 $f_0(\beta=0.74)/\sqrt{2} \simeq 76{\rm\,MeV}$.
 The estimates of the universal low energy scales $\Lambda_{3,4}$ are
 within the bounds of their phenomenological values given in
 \cite{DURR}:
 $\Lambda_3 = 0.6\,(+1.4,-0.4){\rm\,GeV}$,
 $\Lambda_4 = 1.2\,(+0.7,-0.4){\rm\,GeV}$ that is
 $2.3 \leq \Lambda_3/F_0 \leq 23.3$, $9.3 \leq \Lambda_4/F_0 \leq 22.1$.

\vspace*{2em}
\noindent
{\large\bf Acknowledgments}

\noindent
 We thank H.~Perlt for providing us with the perturbative estimates of
 the renormalisation constants of the quark bilinears.
 The computer centers at DESY Hamburg, NIC, DESY Zeuthen, NIC at
 Forschungszentrum J{\"u}lich and HLRN provided the necessary technical
 help and computer resources. 
 This work was supported by the DFG Sonderforschungsbereich/Transregio
 SFB/TR9-03.



\newpage
\appendix
\section{Appendix}
\subsection{Comparison with the fits to plaquette-action data}\label{appendix}

 It is interesting to compare the results obtained from the DBW2 gauge
 action with those presented in \cite{PLAQUETTE} resulting from the
 plaquette gauge action.
 As shown in the previous sections, Chiral Perturbation Theory for Wilson
 lattice fermions (WChPT) offers a natural framework to perform such
 comparison.
 In fact, if NLO WChPT is applicable, the parameters $B_0$, $F_0$ and
 $L_i$ entering equations (\ref{eq55}), (\ref{eq56}) and
 (\ref{eq57}) should already take their physical (continuum) values:
 lattice artefacts are expected to be taken into account by the
 $W$ parameters.
 The latter depend, in general, on the lattice action.

 We remark that, having expressed all quantities ($F_\pi$, $G_\pi$
 and $m_{\pi}$) as functions of $\mcp$, the parameter $W'$
 (see \cite{SHARPE-WU1,SHARPE-WU2}) disappears, and the pion mass can
 apparently go to zero when $\mcp\rightarrow 0$ and $\mu\rightarrow 0$.
 However, one should keep in mind that not all values of $\mcp$ are
 accessible with stable simulation points.
 This parametrization allows to include in the ChPT fit also metastable
 points, where both $m_{\pi}$ and $\mcp$ are lower than it would be
 possible in a stable minimum of the effective potential. 
 Since this is an interesting check, we exploit this possibility and we
 include also metastable points (from \cite{PLAQUETTE}) in the  fit.

 Given the larger amount of data points, we use a different fit procedure
 from the one described in Section \ref{sec4.1}.
 The $\chi^2$ is defined as for the effective variances method
 \cite{OREAR,MARSHALL-BLENCOE,NumRec}, but minimized through the
 {\em Matlab} implementation of the Nelder-Mead Simplex Method.
 The variables $a$ and $\mcp$ are taken as independent variables, and
 $F_\pi$, $G_\pi$ and $m_{\pi}$ as dependent ones.

\begin{figure}
\centering
\begin{minipage}[c]{1.0\linewidth}
\hspace*{0.08\hsize}
\includegraphics[angle=-90,width=0.75\hsize]{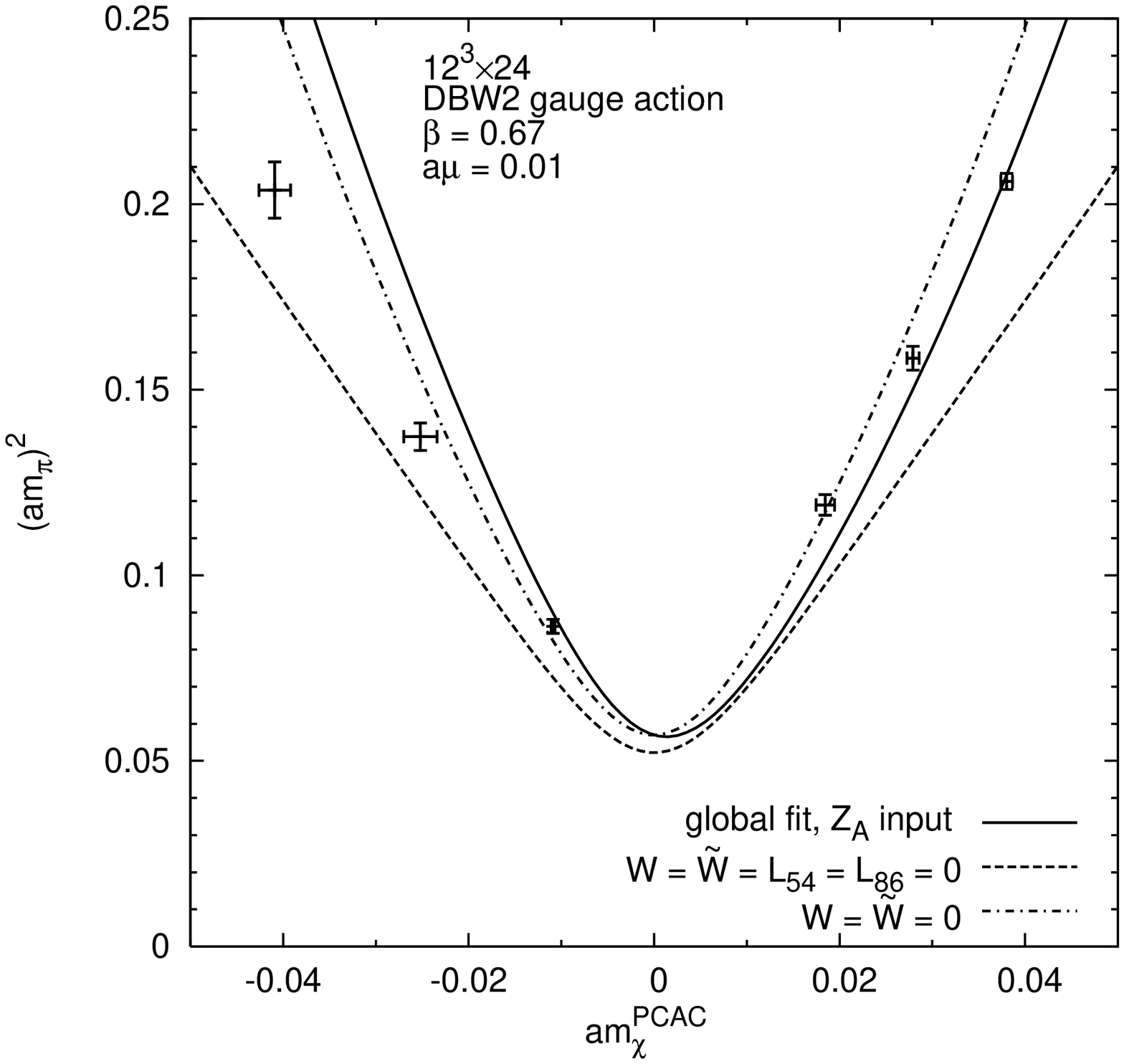}
\end{minipage}
\parbox{0.8\linewidth}{\caption{\label{mpi-fit-067}\em
 The charged pion masses squared as a function of $\amcp$ at
 $a\mu = 0.01$.
 The points represent the data at $\beta=0.67$.
 The solid line displays the global fit with $Z_A$ as input.
 The dashed and dotted lines show the fit with part of the $L$ and $W$
 coefficients set to zero, in order to indicate the size of the NLO
 corrections.}}
\end{figure}

\begin{figure}
\centering
\begin{minipage}[c]{1.0\linewidth}
\hspace*{0.08\hsize}
\includegraphics[angle=-90,width=0.75\hsize]{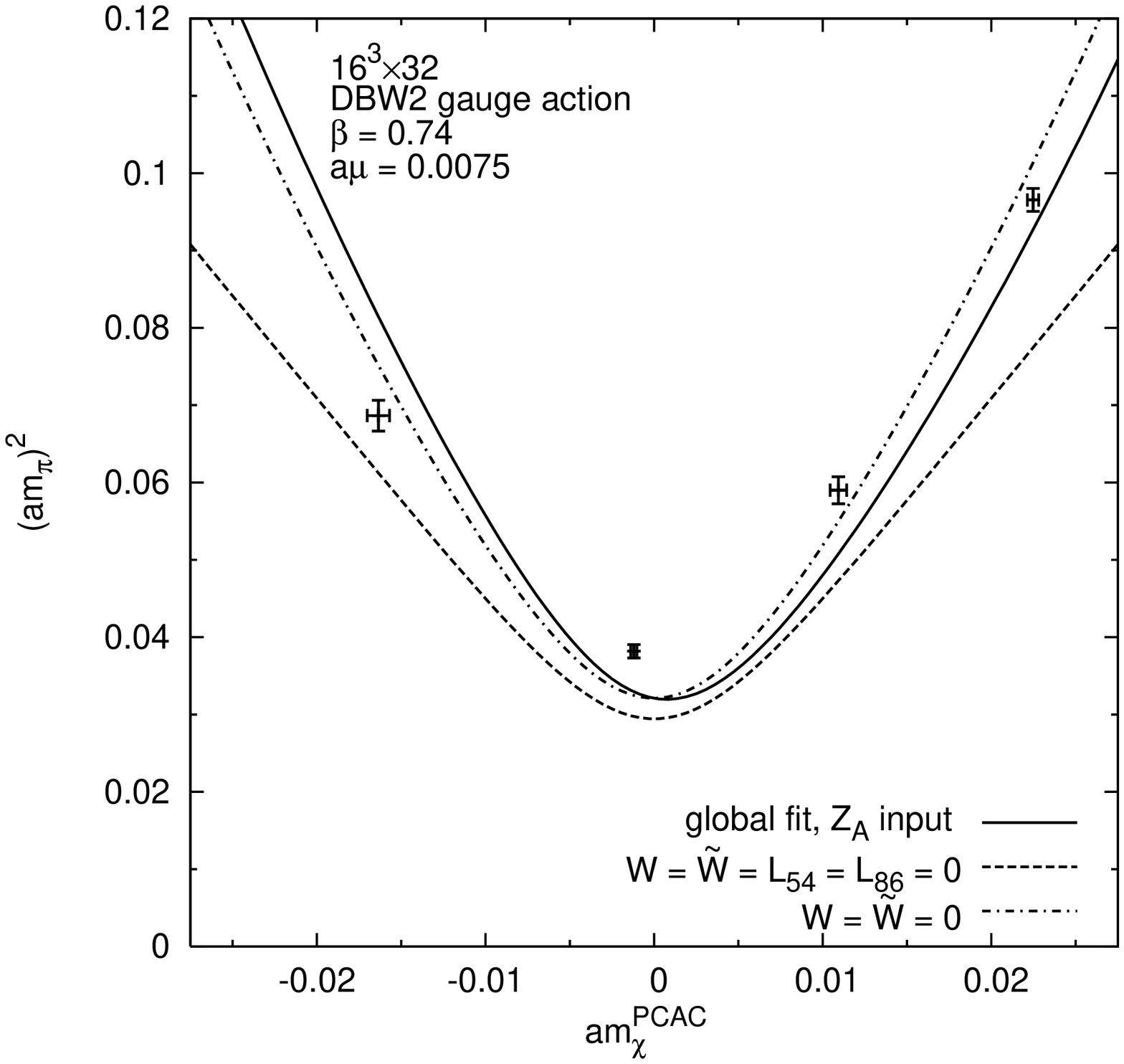}
\end{minipage}
\parbox{0.8\linewidth}{\caption{\label{mpi-fit-074}\em
 The charged pion masses squared as a function of $\amcp$ at
 $a\mu = 0.0075$.
 The points represent the data at $\beta=0.74$.
 The solid line displays the global fit with $Z_A$ as input.
 The dashed and dotted lines show the fit with part of the $L$ and $W$
 coefficients set to zero, in order to indicate the size of the NLO
 corrections.}}
\end{figure}

\begin{figure}
\centering
\begin{minipage}[c]{1.0\linewidth}
\hspace*{0.08\hsize}
\includegraphics[angle=-90,width=0.75\hsize]{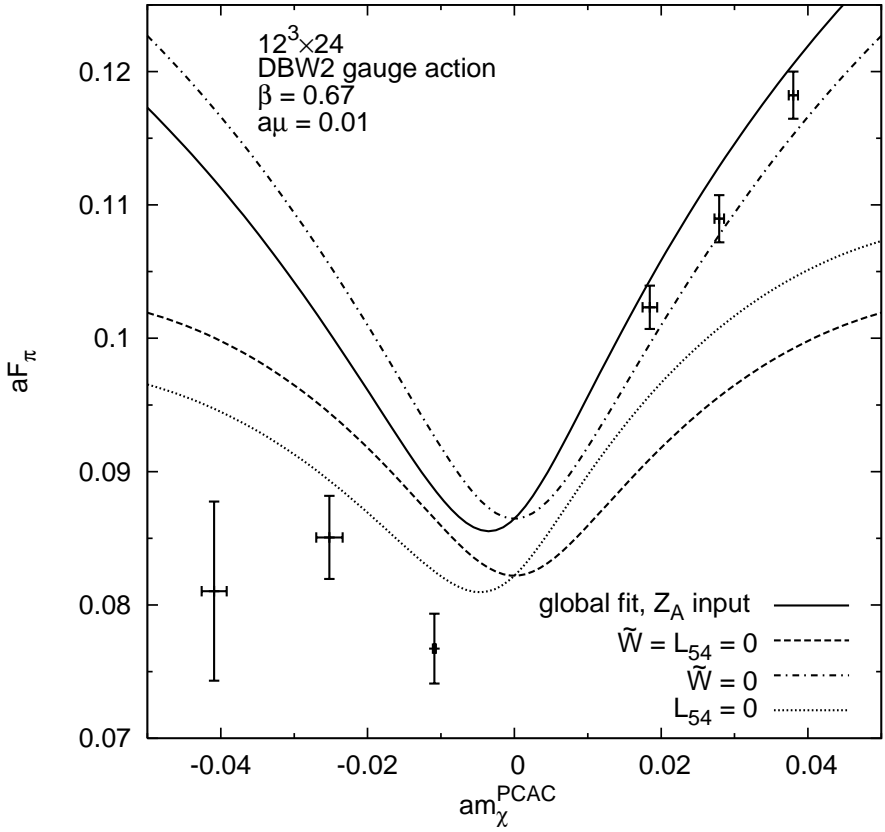}
\end{minipage}
\parbox{0.8\linewidth}{\caption{\label{fpi-fit}\em
 The pion decay constant $aF_\pi$ as a function of $\amcp$ at
 $\beta = 0.67$, $a\mu=0.01$.
 The solid line displays the global fit with $Z_A$ as input.
 The dashed and dotted lines show the fit with part of the $L$ and $W$
 coefficients set to zero, in order to indicate the size of the NLO
 corrections.}}
\end{figure}

\begin{figure}
\centering
\begin{minipage}[c]{1.0\linewidth}
\hspace*{0.08\hsize}
\includegraphics[angle=-90,width=0.75\hsize]{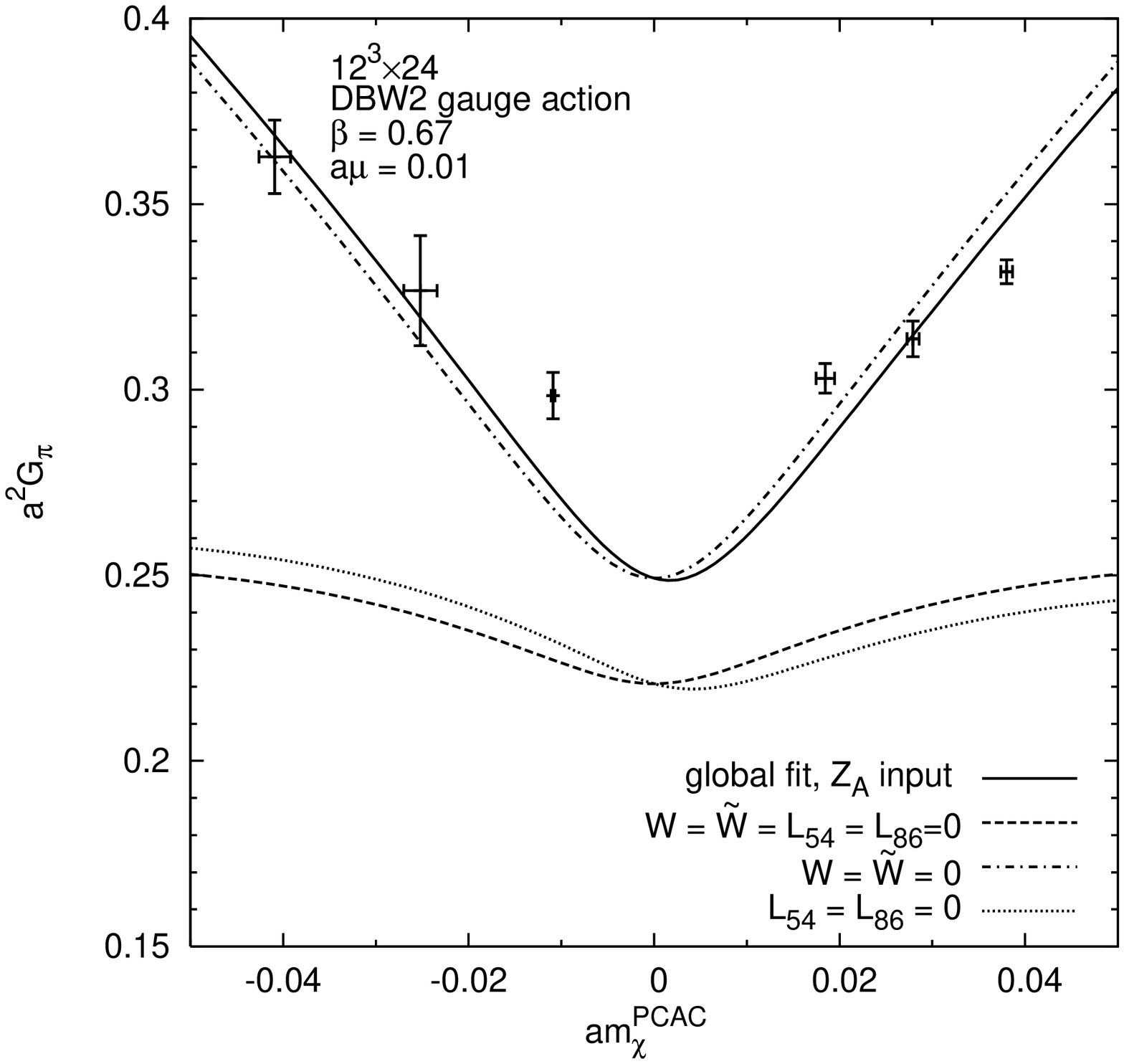}
\end{minipage}
\parbox{0.8\linewidth}{\caption{\label{gpi-fit}\em
 The pseudoscalar matrix element $a^2G_\pi$ as a function of $\amcp$ at
 $\beta = 0.67$, $a\mu = 0.0075$.
 The solid line displays the global fit with $Z_A$ as input.
 The dashed and dotted lines show the fit with part of the $L$ and $W$
 coefficients set to zero, in order to indicate the size of the NLO
 corrections.}}
\end{figure}

\begin{figure}
\centering
\begin{minipage}[c]{1.0\linewidth}
\hspace{0.08\hsize}
\includegraphics[angle=-90,width=0.75\hsize]
 {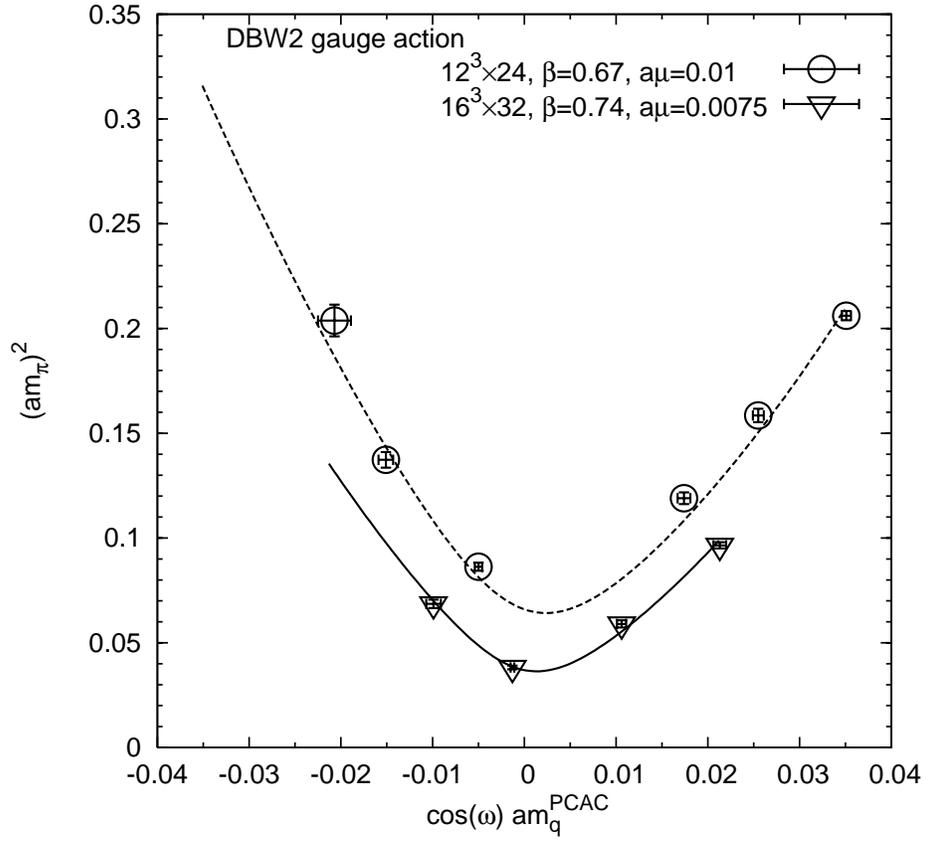}
\end{minipage}
\parbox{0.8\linewidth}{
\caption{\label{fig_mpi_dbw} 
 Fit of the charged pion mass squared from DBW2 data at non-zero $a\mu$
 as described in Sec.~\ref{appendix}.
 Upper (lower) curve belongs to $\beta=0.67$ ($\beta=0.74$).}}
\end{figure}

\begin{figure}
\centering
\begin{minipage}[c]{1.0\linewidth}
\hspace{0.08\hsize}
\includegraphics[angle=-90,width=0.75\hsize]
 {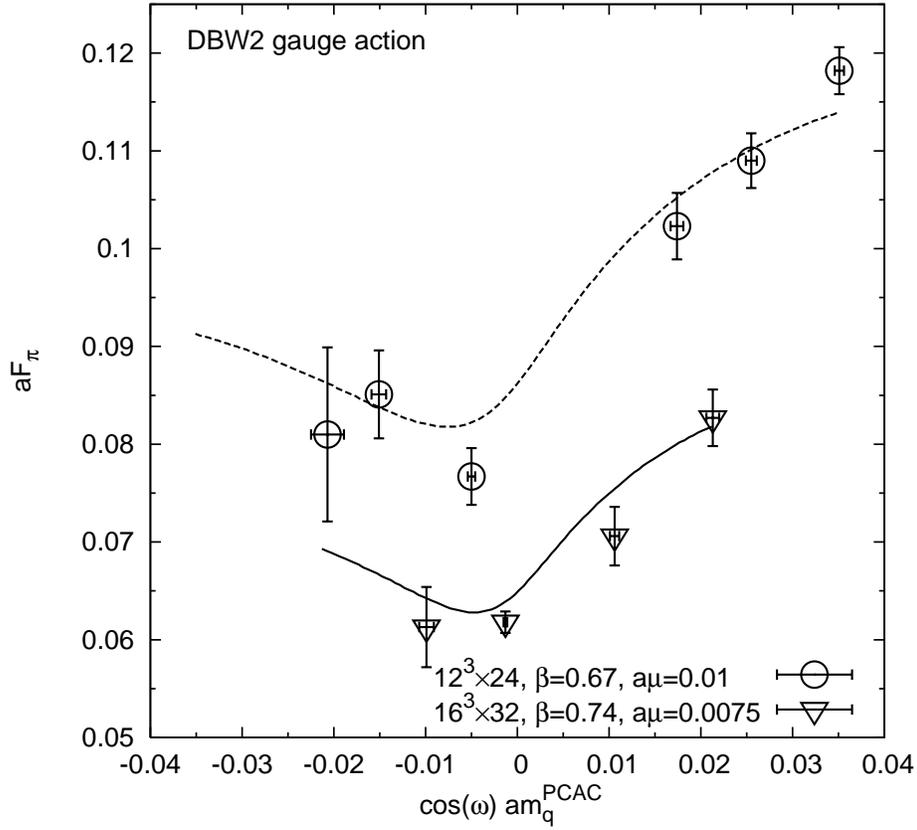}
\end{minipage}
\parbox{0.8\linewidth}{
\caption{\label{fig_fpi_dbw} 
 Fit of the pion decay constant $aF_\pi$ from DBW2 data at non-zero
 $a\mu$ as described in Sec.~\ref{appendix}.
 Upper (lower) curve belongs to $\beta=0.67$ ($\beta=0.74$).}}
\end{figure}

\begin{figure}
\centering
\begin{minipage}[c]{1.0\linewidth}
\hspace{0.08\hsize}
\includegraphics[angle=-90,width=0.75\hsize]
 {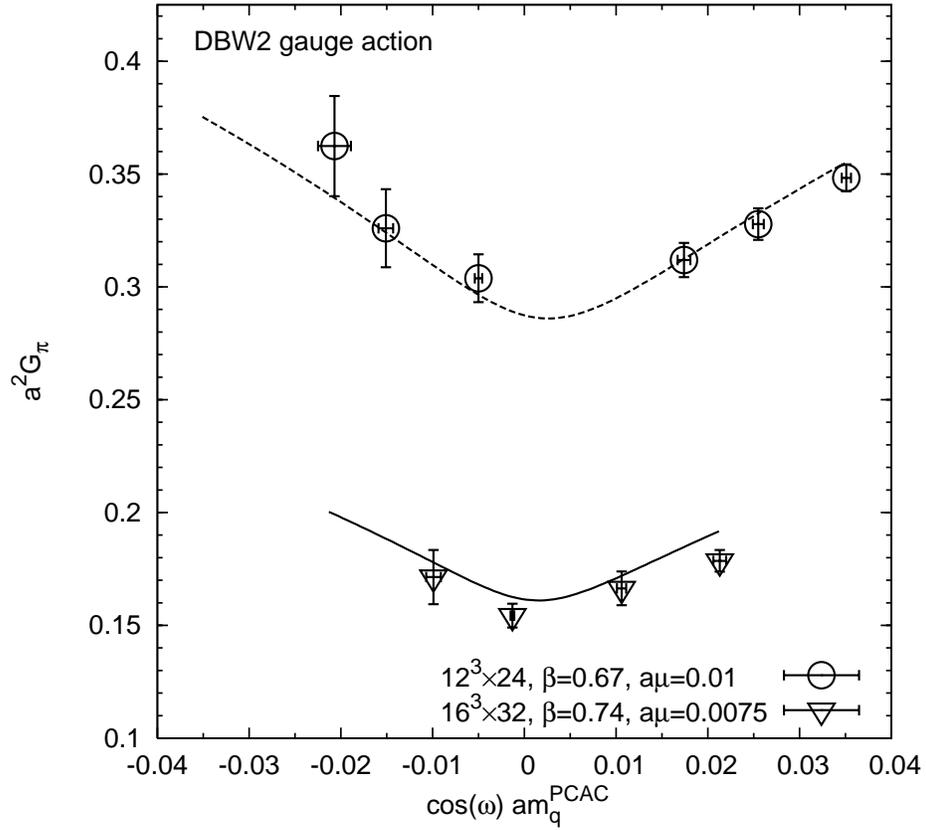}
\end{minipage}
\parbox{0.8\linewidth}{
\caption{\label{fig_gpi_dbw}
 Fit of $a^2 G_\pi$ from DBW2 data at non-zero $a\mu$ as described in
 Sec.~\ref{appendix}.
 Upper (lower) curve belongs to $\beta=0.67$ ($\beta=0.74$).}}
\end{figure}

 Besides using a fitting procedure different from the one in the
 previous Sections \ref{sec4.1}-\ref{sec4.2}, our fits to the
 plaquette gauge action data are restricted to data points with non-zero
 twisted mass ($a\mu>0$) only.
 We also tried to use different independent variables instead
 of $\amcp$, which correspond to different possible definitions of the
 untwisted component of the quark mass.
 It turned out that the fit quality is improving if one considers
 $\amcpb$ defined in (\ref{eq37}).
 The difference implied by these changes compared to the analysis in
 Sections \ref{sec4.1}-\ref{sec4.2} -- i.e.~different fitting
 procedure, restricting the fit to $a\mu>0$ and using $\amcpb$ --
 is illustrated by Figures~\ref{fig_mpi_dbw}, \ref{fig_fpi_dbw} and
 \ref{fig_gpi_dbw} which have to be compared to
 Figures~\ref{mpi-fit-067}, \ref{mpi-fit-074}, \ref{gpi-fit} and
 \ref{fpi-fit}, respectively.

 A consequence of considering $\amcpb$ instead of $\amcp$ is that
 $Z_A$ enters only indirectly -- through the determination of $\omega$
 -- therefore we do not need to fit them.
 As said before, the $Z_P$ is included in the $B$ factor.
 However, when comparing different lattice spacings and different
 actions, we must allow a $\beta$ dependent $Z_P$.
 In practice we choose a reference $\beta$ (corresponding to the
 smallest $a$ which appears in the fit) and we fit a correction to
 $Z_P$ for each different $a$.
 These are not given in the table, but they are always between $0.95$
 and $1.35$.

 We summarize our results for the plaquette gauge action data in
 Table~\ref{fit_resu_plq}.
 No statistical errors are quoted, since the systematic errors
 dominate, as the comparison of the results from the different fit
 setups shows.
 We perform fits including all data (top part of
 Table~\ref{fit_resu_plq}) or only data at positive mass (bottom part
 of Table~\ref{fit_resu_plq}). 
 In this second case the $W$ parameters are set to zero.

\begin{figure}
\centering
\begin{minipage}[c]{1.0\linewidth}
\hspace{0.08\hsize}
\includegraphics[angle=-90,width=0.75\hsize]
 {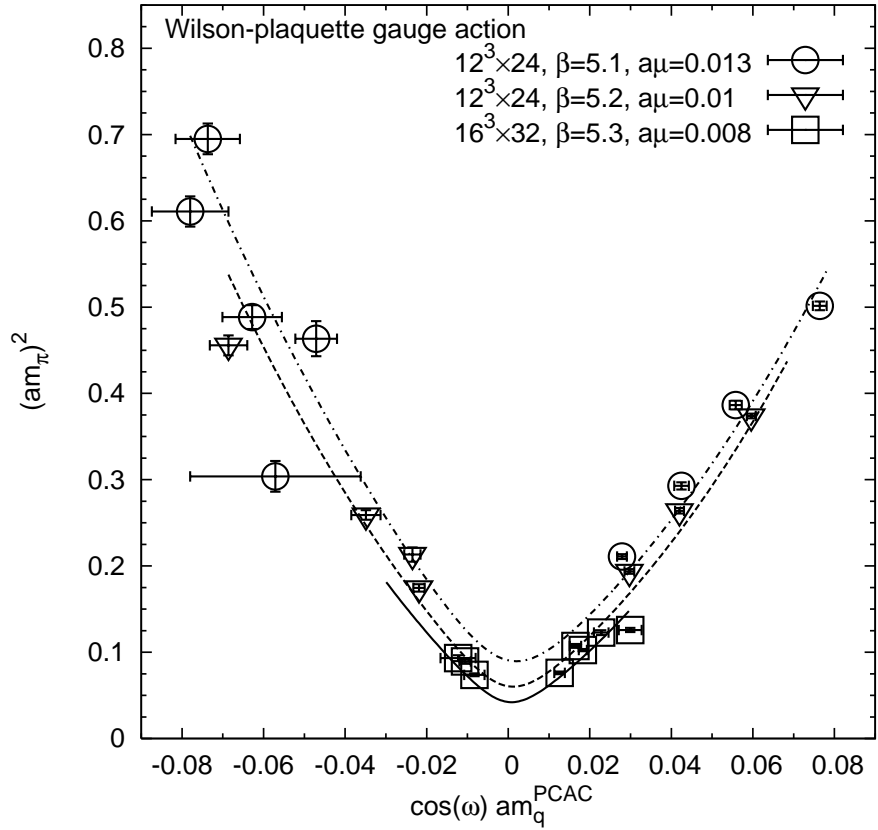}
\end{minipage}
\parbox{0.8\linewidth}{
\caption{\label{fig_mpi_plq}
 Fit of the charged pion mass squared from plaquette data at non-zero
 $a\mu$.
 Upper, intermediate and lower curves refer to $\beta=5.1$,
 $\beta=5.2$ and $\beta=5.3$, respectively.}}
\end{figure}

\begin{figure}
\centering
\begin{minipage}[c]{1.0\linewidth}
\hspace{0.08\hsize}
\includegraphics[angle=-90,width=0.75\hsize]
 {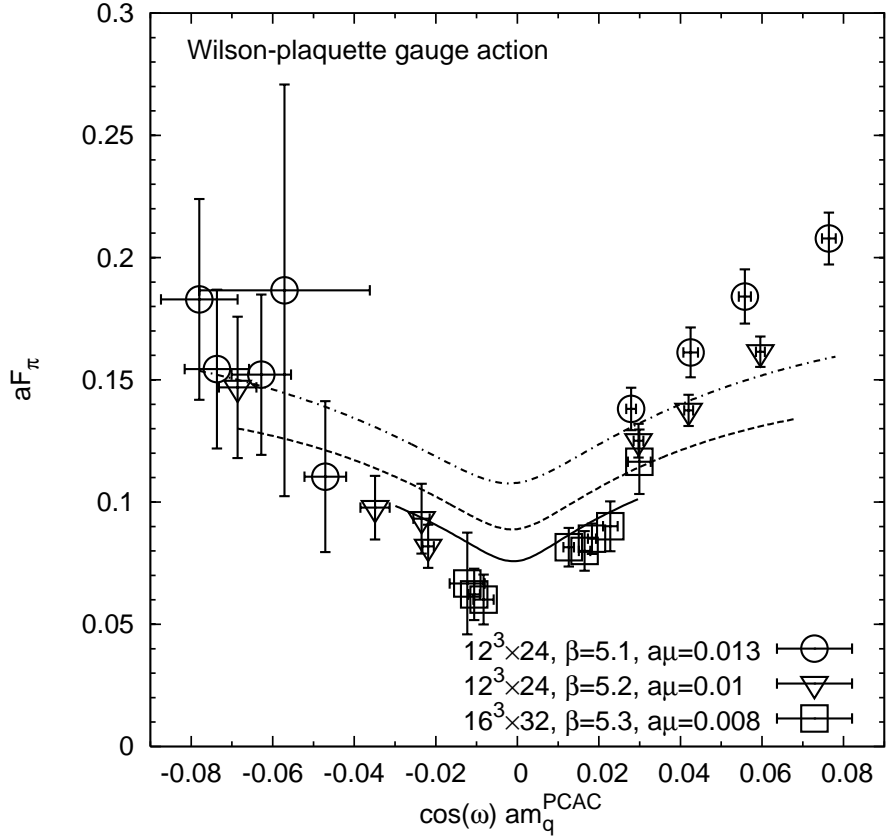}
\end{minipage}
\parbox{0.8\linewidth}{
\caption{\label{fig_fpi_plq}
 Fit of the pion decay constant $aF_\pi$ from plaquette data at non-zero
 $a\mu$.
 Upper, intermediate and lower curves refer to $\beta=5.1$,
 $\beta=5.2$ and $\beta=5.3$, respectively.}}
\end{figure}

\begin{figure}
\centering
\begin{minipage}[c]{1.0\linewidth}
\hspace{0.08\hsize}
\includegraphics[angle=-90,width=0.75\hsize]
 {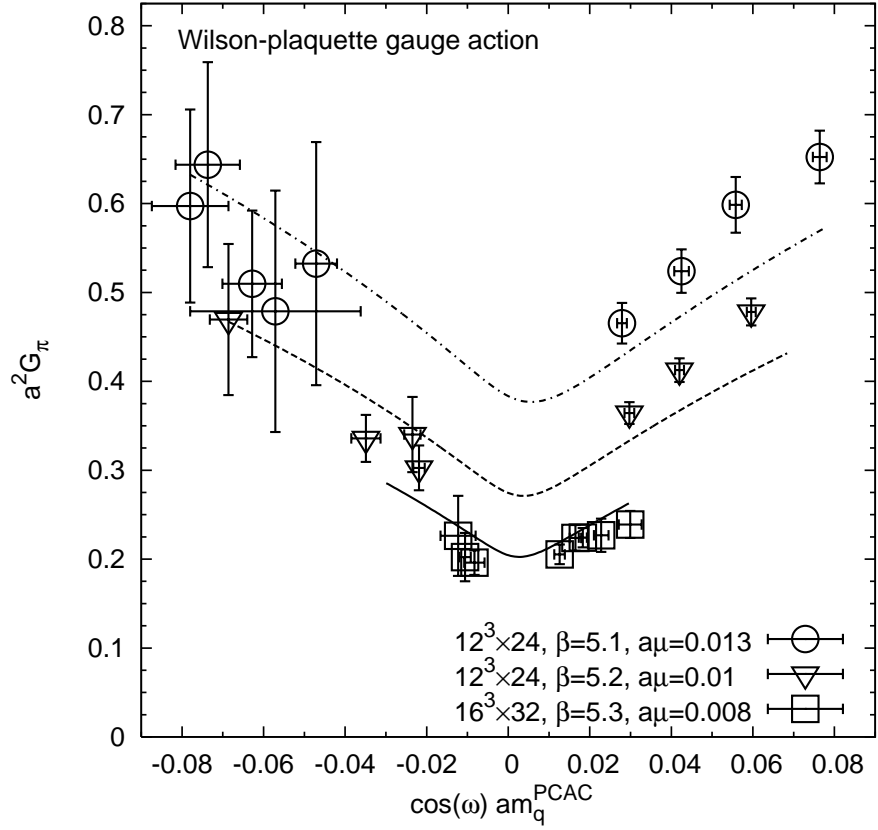}
\end{minipage}
\parbox{0.8\linewidth}{
\caption{\label{fig_gpi_plq}
 Fit of $a^2 G_\pi$ from plaquette data at non-zero $a\mu$.
 Upper, intermediate and lower curves refer to $\beta=5.1$,
 $\beta=5.2$ and $\beta=5.3$, respectively.}}
\end{figure}

 In Figures \ref{fig_mpi_plq}, \ref{fig_fpi_plq} and \ref{fig_gpi_plq}
 the fits of the plaquette gauge action data are presented.
 Similarly to the DBW2 fits, the $W$'s are very unstable, depending on
 the chosen subset of data, and in general they are consistent with zero
 within errors. 
 The physical combinations $L_{54}$ and $L_{86}$ are consistent with
 the values obtained by the DBW2 gauge action.

 We also performed fits of all data and imposing $W=\tilde{W}=0$.
 The values of the physical quantities are still reasonable, however
 the curves fit the data worse.
 We have also attempted fits where all the NLO parameters are set to
 zero ($L_i=W=\tilde{W}=0$), or where only lattice artefacts are
 included ($L_i=0$).
 Both these assumptions result in very poor fits, essentially because
 they cannot reproduce the curvature in $F_\pi$ and $G_\pi$.

\begin{table}
\begin{center}
{\Large\bf Tables}
\end{center}
\vspace*{1em}
\begin{center}
\parbox{0.8\linewidth}{\caption{\label{tab_run}\em
 Run parameters: the gauge coupling ($\beta$), the twisted mass
 in lattice units ($a\mu$), the hopping parameter ($\kappa$)
 and the lattice size.
 The last column shows the number of gauge configurations used
 in the data analysis.}}
\end{center}
\begin{center}
\renewcommand{\arraystretch}{1.2}
\begin{tabular}{|c*{5}{|c}|}
 \hline
 run & $\beta$ & $a\mu$ & $\kappa$ & $L^3\times T$ & $N_{\rm conf}$ \\
 \hline
 ($a$) & 0.67 & 0 & 0.1650 & $12^3\times24$ & 4514 \\
 \hline
 ($b$) & 0.67 & 0 & 0.1655 & $12^3\times24$ & 2590 \\
 \hline
 ($c$) & 0.67 & 0 & 0.1660 & $12^3\times24$ & 2589 \\
 \hline
 ($d$) & 0.67 & 0 & 0.1665 & $12^3\times24$ & 1721 \\
 \hline
 \hline
 ($a'$) & 0.67 & 0.01 & 0.1650 & $12^3\times24$ & 600 \\
 \hline
 ($b'$) & 0.67 & 0.01 & 0.1655 & $12^3\times24$ & 620 \\
 \hline
 ($c'$) & 0.67 & 0.01 & 0.1660 & $12^3\times24$ & 509 \\
 \hline
 ($d'$) & 0.67 & 0.01 & 0.1665 & $12^3\times24$ & 570 \\
 \hline
 ($e'$) & 0.67 & 0.01 & 0.1670 & $12^3\times24$ & 584 \\
 \hline
 ($f'$) & 0.67 & 0.01 & 0.1675 & $12^3\times24$ & 499 \\
 \hline
 ($g'$) & 0.67 & 0.01 & 0.1680 & $12^3\times24$ & 606 \\
 \hline
 \hline
 ($A$) & 0.74 & 0 & 0.1580 & $16^3\times32$ & 1319 \\
 \hline
 ($B$) & 0.74 & 0 & 0.1585 & $16^3\times32$ & 419 \\
 \hline
 \hline
 ($A'$) & 0.74 & 0.0075 & 0.1580 & $16^3\times32$ & 430 \\
 \hline
 ($B'$) & 0.74 & 0.0075 & 0.1585 & $16^3\times32$ & 296 \\
 \hline
 ($C'$) & 0.74 & 0.0075 & 0.1590 & $16^3\times32$ & 353 \\
 \hline
 ($D'$) & 0.74 & 0.0075 & 0.1595 & $16^3\times32$ & 352 \\
 \hline
\end{tabular}
\end{center}
\end{table}

\begin{table}
\begin{center}
\parbox{0.8\linewidth}{\caption{\label{tab_one}\em
 The results for the scale parameter ($r_0/a$), the pseudoscalar
 (``pion'') mass $am_\pi$ and the vector-meson (``$\rho$-meson'') mass
 $am_\rho$.}}
\end{center}
\begin{center}
\renewcommand{\arraystretch}{1.2}
\begin{tabular}{*{7}{|l}|}
  \hline 
  run & \,\,\, $r_0/a$ & \,\,\,\, $am_{\pi}$ & 
  \,\,\,\, $am_{\rho}$ & \,\,\, $m_{\pi}/m_{\rho}$ & 
  \,\,\, $r_0m_{\pi}$ & \, $(r_0m_{\pi})^2$           \\
%
%
  \hline
  ($a$) & 2.305(36) & 0.4468(30) & 0.7025(44) & 0.6359(51) &
  1.030(19) & 1.061(38) \\
  \hline
  ($b$) & 2.391(56) & 0.4085(55) & 0.7007(79) & 0.5831(66) &
  0.977(23) & 0.954(44)  \\
  \hline
  ($c$) & 2.351(27) & 0.3619(27) & 0.629(10) & 0.5747(84) &
  0.850(11) & 0.724(19) \\
  \hline
  ($d$) & 2.652(38) & 0.235(12) & 0.595(22) & 0.396(18) &
  0.623(30) & 0.389(37) \\
  \hline
%
%
  \hline
  ($a'$) & 2.347(26) & 0.4540(24) & 0.7026(46) & 0.6461(47) &
  1.065(12) & 1.135(25) \\
  \hline
  ($b'$) & 2.415(24) & 0.3981(40) & 0.6808(66) & 0.5847(61) &
  0.9618(25) & 0.925(18)  \\
  \hline
  ($c'$) & 2.503(29) & 0.3449(40) & 0.662(11) & 0.520(10) &
  0.863(11) & 0.745(18) \\
  \hline
  ($d'$) & 2.867(29) & 0.2793(26) & 0.654(45) & 0.426(30) &
  0.801(16) & 0.641(26)  \\
  \hline
  ($e'$) & 3.127(31) & 0.2937(32) & 0.807(64) & 0.363(29) &
  0.918(14) & 0.844(25) \\
  \hline
  ($f'$) & 3.279(36) & 0.3706(50) & 0.913(72) & 0.403(33) &
  1.215(23) & 1.477(57) \\
  \hline
  ($g'$) & 3.261(31) & 0.4514(84) & 1.013(82) & 0.444(36) &
  1.472(30) & 2.168(88) \\
  \hline
%
%
  \hline
  ($A$) & 3.563(33) & 0.3038(15) & 0.5256(37) & 0.5780(41) &
  1.082(11) & 1.172(23) \\
  \hline
  ($B$) & 3.741(90) & 0.2250(29) & 0.491(14) & 0.457(13) &
  0.843(22) & 0.711(36) \\
  \hline
%
%
  \hline
  ($A'$) & 3.467(51) & 0.3107(24) & 0.5354(71) & 0.5803(78)
  & 1.077(17) & 1.161(36) \\
  \hline
  ($B'$) & 3.78(10) & 0.2429(36) & 0.537(21) & 0.451(18)
  & 0.920(25) & 0.846(46)  \\
  \hline
  ($C'$) & 3.87(10) & 0.1954(22) & 0.57(14) & 0.337(79)
  & 0.756(31) & 0.572(48) \\
  \hline
  ($D'$) & 4.148(65) & 0.2620(38) & 0.639(73) & 0.409(48)
  & 1.086(24) & 1.181(52) \\
  \hline
\end{tabular}
\end{center}
\end{table}

\begin{table}
\begin{center}
\parbox{0.8\linewidth}{\caption{\label{tab_two}\em
 The results for the PCAC quark mass ($\amcp$) and pseudoscalar
 (``pion'') decay constant ($af_{\chi\pi}$).}}
\end{center}
\begin{center}
\renewcommand{\arraystretch}{1.2}
\begin{tabular}{*{5}{|l}|}
  \hline
  run & \,\, $\amcp$ & \,\, $r_0\mcp$ & 
  \,\,\,\,\, $af_{\chi\pi}$ & \,\,\,\,\, $r_0f_{\chi\pi}$  \\
%
%
  \hline
  ($a$) & 0.03884(22) & 0.0895(14) & 0.18567(90) & 0.4279(62) \\
  \hline
  ($b$) & 0.03224(71) & 0.0771(18) & 0.1798(17) & 0.4301(98) \\
  \hline
  ($c$) & 0.02247(80) & 0.0528(20) & 0.1553(27) & 0.3653(75)  \\
  \hline
  ($d$) & 0.00972(43) & 0.0258(11) & 0.1369(65) & 0.363(18)  \\
  \hline 
%
%
  \hline
  ($a'$) & 0.03801(63) & 0.0892(16) & 0.05774(88) & 0.1355(25) \\
  \hline
  ($b'$) & 0.02791(65) & 0.0674(16) & 0.0520(11) & 0.1257(28) \\
  \hline
  ($c'$) & 0.01846(99) & 0.0462(22) & 0.0442(20) & 0.1107(44) \\
  \hline
  ($d'$) & 0.00505(82) & 0.0145(22) & 0.0174(26) & 0.0499(75) \\
  \hline
  ($e'$) & -0.0109(2) & -0.0341(37) & -0.0354(37) & -0.110(12)  \\
  \hline
  ($f'$) & -0.0252(18) & -0.0829(62) & -0.0562(44) & -0.184(15)  \\
  \hline 
  ($g'$) & -0.0409(17) & -0.1336(56) & -0.0683(30) & -0.2229(98) \\
  \hline
%
%
  \hline
  ($A$) & 0.02313(23) & 0.0824(10) & 0.1243(12) & 0.4429(58) \\
  \hline
  ($B$) & 0.01251(43) & 0.0469(24) & 0.1124(37) & 0.420(22) \\
  \hline
%
%
  \hline
  ($A'$)& 0.02247(33) & 0.0779(16) & 0.03645(60) & 0.1264(28) \\
  \hline
  ($B'$)& 0.01093(49) & 0.0414(21) & 0.0266(12) & 0.1007(46) \\
  \hline
  ($C'$)& -0.00120(18) & -0.0046(29) & -0.0043(18) & -0.016(11)  \\
  \hline
  ($D'$)& -0.01635(66) & -0.06783(29) & -0.0361(16) & -0.1500(71)  \\
  \hline 
\end{tabular}
\end{center}
\end{table}

\begin{table}
\centering
\begin{center}
\parbox{0.8\linewidth}{\caption{\label{tab_omega}\em
The twist angles $\omega$, $\omega_V$ and $\omega_A$,
as defined in Eqs.~(\ref{eq08}), (\ref{eq09}) and
(\ref{eq12}), determined by Eqs.~(\ref{eq22}), (\ref{eq23}) and (\ref{eq16}). 
}}
\end{center}
\begin{center}
\begin{tabular}{|l|l|l|l|l|l|}
\hline
\multicolumn{1}{|c|}{$\beta$} & 
\multicolumn{1}{|c|}{$a\mu$} & 
\multicolumn{1}{|c|}{$\kappa$} & 
\multicolumn{1}{|c|}{$\omega_V/\pi$} &
\multicolumn{1}{c}{$\omega_A/\pi$} & 
\multicolumn{1}{|c|}{$\omega/\pi$} 
\\ \hline
0.67 & $1.0\, \cdot 10^{-2}$ & 0.1650  & 0.1352(13) & 0.0564(17)  & 0.0883(13)   \\
\hline
0.67 & $1.0\, \cdot 10^{-2}$ & 0.1655  & 0.1772(29) & 0.0771(27)  & 0.1190(25)   \\
\hline
0.67 & $1.0\, \cdot 10^{-2}$ & 0.1660  & 0.2412(62) & 0.1069(41)  & 0.1661(54)   \\
\hline
0.67 & $1.0\, \cdot 10^{-2}$ & 0.1665  & 0.411(12)  & 0.229(17)   & 0.334(17)    \\
\hline
0.67 & $1.0\, \cdot 10^{-2}$ & 0.1670  & 0.678(12)  & 0.622(16)   & 0.647(11)    \\
\hline
0.67 & $1.0\, \cdot 10^{-2}$ & 0.1675  & 0.8053(86) & 0.826(13)   & 0.8137(80)   \\
\hline
0.67 & $1.0\, \cdot 10^{-2}$ & 0.1680  & 0.8709(43) & 0.843(23)   & 0.857(11)    \\
\hline
\hline
0.74 & $7.5\, \cdot 10^{-3}$ & 0.1580  & 0.1542(26) & 0.0722(38)  & 0.1076(31)   \\
\hline
0.74 & $7.5\, \cdot 10^{-3}$ & 0.1585  & 0.2613(66) & 0.1393(96)  & 0.1963(77)   \\
\hline
0.74 & $7.5\, \cdot 10^{-3}$ & 0.1590  & 0.532(12)  & 0.582(37)   & 0.5544(92)   \\
\hline
0.74 & $7.5\, \cdot 10^{-3}$ & 0.1597  & 0.7966(49) & 0.790(15)   & 0.794(12)    \\
\hline
\end{tabular}
\end{center}
\end{table}

\begin{table}
\centering
\begin{center}
\parbox{0.8\linewidth}{\caption{\label{tab_kcr}\em
 Determination of $\mu_{\kappa cr}$ by requiring $\omega=\pi/2$, 
 $\mu_{\kappa cr}(\omega_V)$, or $\mcp=0$, $\mu_{\kappa cr}(\mcp)$.
 The plus and minus signs indicate extrapolations from positive or negative 
 untwisted quark masses $\mcp$, {\rm avg} denotes the average.
}}
\end{center}
\begin{center}
\begin{tabular}{|l|l|c|l|l|l|}
\hline
\multicolumn{1}{|c|}{$\beta$} &
\multicolumn{1}{|c|}{$a\mu$} &
\multicolumn{1}{|c|}{sign} &
\multicolumn{1}{|c|}{$\mu_{\kappa cr}(\omega_V)$} &
\multicolumn{1}{|c|}{$\mu_{\kappa cr}(\mcp)$}&
\multicolumn{1}{|c|}{$Z_{oV}$}
\\ \hline
0.67 & $1.0\, \cdot 10^{-2}$ & $+$  & 2.99800(9)  & 2.99839(12) &  1.438(33)      \\
0.67 & $1.0\, \cdot 10^{-2}$ & $-$  & 3.00059(13) & 3.00043(17) &  1.065(61)      \\
0.67 & $1.0\, \cdot 10^{-2}$ & avg  & 2.99930(11) & 2.99941(15) &  1.251(47)      \\
\hline
0.74 & $7.5\, \cdot 10^{-3}$& $+$  & 3.145528(52)  & 3.145645(22)   &  1.328(36)  \\
0.74 & $7.5\, \cdot 10^{-3}$& $-$  & 3.145441(52)  & 3.145435(21)   &  1.055(49)  \\
0.74 & $7.5\, \cdot 10^{-3}$& avg  & 3.145484(52)  & 3.145540(22)   &  1.191(42)  \\
\hline
\end{tabular}
\end{center}
\end{table}

\begin{table}
\centering
\begin{center}
\parbox{0.8\linewidth}{\caption{\label{tab_zfac}\em
 Renormalisation constants of the vector and axialvector currents.
 The ratio $Z_A/Z_V$ is determined from the analysis of the twist angles,
 cf. Eq.~(\ref{eq16});
 two different determinations of the vector current $Z_V$ are reported:
 $Z_V^{(1)}$ from Eq.~(\ref{eq30}) and $Z_V^{(2)}$ from Eq.~(\ref{eq32});
 the renormalisation constant of the axialvector current is derived by
 combining the results for $Z_A/Z_V$ and $Z_V^{(1)}$.
}}
\end{center}
\begin{center}
\begin{tabular}{|l|l|l|l|l|l|l|}
\hline
\multicolumn{1}{|c|}{$\beta$} & 
\multicolumn{1}{|c|}{$a\mu$} & 
\multicolumn{1}{|c|}{$\kappa$} & 
\multicolumn{1}{|c|}{$Z_A/Z_V$} & 
\multicolumn{1}{|c|}{$Z_V^{(1)}$} &
\multicolumn{1}{|c|}{$Z_V^{(2)}$} &  
\multicolumn{1}{|c|}{$Z_A$}
\\ \hline
0.67 & $1.0\, \cdot 10^{-2}$ & 0.1650  & 1.589(26)  & 0.5910(13)  & 0.5810(16) & 0.939(15) \\
\hline
0.67 & $1.0\, \cdot 10^{-2}$ & 0.1655  & 1.587(28)  & 0.5813(11)  & 0.5761(25) & 0.923(16) \\
\hline
0.67 & $1.0\, \cdot 10^{-2}$ & 0.1660  & 1.649(28)  & 0.5766(12)  & 0.5708(38) & 0.951(16) \\
\hline
0.67 & $1.0\, \cdot 10^{-2}$ & 0.1665  & 1.979(68)  & 0.5689(10)  & 0.5657(39) & 1.126(39) \\
\hline
0.67 & $1.0\, \cdot 10^{-2}$ & 0.1670  & 0.815(58)  & 0.5705(14)  & 0.5666(46) & 0.465(33) \\
\hline
0.67 & $1.0\, \cdot 10^{-2}$ & 0.1675  & 1.087(47)  & 0.5716(32)  & 0.5688(38) & 0.623(27) \\
\hline
0.67 & $1.0\, \cdot 10^{-2}$ & 0.1680  & 0.894(78)  & 0.5851(33)  & 0.5754(43) & 0.518(46) \\
\hline
\hline
0.74 & $7.5\, \cdot 10^{-3}$ & 0.1580  & 1.508(35)  & 0.6379(12)  & 0.6315(32) & 0.963(22) \\
\hline
0.74 & $7.5\, \cdot 10^{-3}$ & 0.1585  & 1.515(59)  & 0.6294(11)  & 0.6294(38) & 0.953(37) \\
\hline
0.74 & $7.5\, \cdot 10^{-3}$ & 0.1590  & 1.65(45)   & 0.62595(95) & 0.6241(38) & 1.04(28) \\
\hline
0.74 & $7.5\, \cdot 10^{-3}$ & 0.1597  & 0.972(73)  & 0.6291(25)  & 0.6242(40) & 0.612(46) \\
\hline
\end{tabular}
\end{center}
\end{table}

\begin{table}
\centering
\begin{center}
\parbox{0.8\linewidth}{\caption{\label{tab_zfac_ft}\em
 Full twist extrapolations for $Z_V$, $Z_A$ and the ratio $Z_A/Z_V$ (see
 text for explanations) with comparison with 1-loop perturbative
 estimates (PT) and tadpole-improved perturbative estimates (TI)\cite{HP}.
 The ratio $Z_P/Z_S$ is also reported, determined from $Z_{oV}$ 
 (see Eqs.~(\ref{eq44}), (\ref{eq45})).
}}
\end{center}
\begin{center}
\begin{tabular}{|l|l|c|c|l|l|l|}\hline
     \multicolumn{1}{|c|}{$\beta$} &
     \multicolumn{1}{|c|}{$a\mu$} &
     \multicolumn{1}{|c|}{Sign} &
     \multicolumn{1}{|c|}{Op.} &
     \multicolumn{1}{|c|}{$Z$} &
     \multicolumn{1}{|c|}{$Z$(PT)}&
     \multicolumn{1}{|c|}{$Z$(TI)}
    \\
\hline
0.67 & $1.0\, \cdot 10^{-2}$ &+& $V$ & 0.5650(11)  &0.6089 &0.6531              \\
0.67 & $1.0\, \cdot 10^{-2}$ &-& $V$ & 0.5673(19)  &0.6089 &0.6531              \\
\hline
0.74 & $7.5\, \cdot 10^{-3}$ &+& $V$ & 0.6217(23)  &0.6459 &0.6892              \\
0.74 & $7.5\, \cdot 10^{-3}$ &-& $V$ & 0.6257(10)  &0.6459 &0.6892              \\
\hline
\hline
0.67 & $1.0\, \cdot 10^{-2}$ &+& $A$ & 0.952(30)   &0.7219 &0.7176             \\
0.67 & $1.0\, \cdot 10^{-2}$ &-& $A$ & 0.49(4)     &0.7219 &0.7176             \\
\hline
0.74 & $7.5\, \cdot 10^{-3}$ &+& $A$ & 0.944(74)   &0.7482 &0.7735             \\
0.74 & $7.5\, \cdot 10^{-3}$ &-& $A$ & 0.612(46)   &0.7482 &0.7735             \\
\hline
\hline
0.67 & $1.0\, \cdot 10^{-2}$ &+& $A/V$ & 1.683(52)   &1.1130 &0.9696             \\
0.67 & $1.0\, \cdot 10^{-2}$ &-& $A/V$ & 0.867(70)   &1.1130 &0.9696             \\
\hline
0.74 & $7.5\, \cdot 10^{-3}$ &+& $A/V$ & 1.52(12)    &1.1023 &0.9747             \\
0.74 & $7.5\, \cdot 10^{-3}$ &-& $A/V$ & 0.972(73)   &1.1023 &0.9747             \\
\hline
\hline
0.67 & $1.0\, \cdot 10^{-2}$ &+& $P/S$ & 1.17(6)     &0.8157 &0.9407            \\
0.67 & $1.0\, \cdot 10^{-2}$ &-& $P/S$ & 0.81(11)    &0.8157 &0.9407             \\
\hline
0.74 & $7.5\, \cdot 10^{-3}$ &+& $P/S$ & 1.14(12)    &0.8302 &0.9444             \\
0.74 & $7.5\, \cdot 10^{-3}$ &-& $P/S$ & 0.92(10)    &0.8302 &0.9444             \\
\hline
\end{tabular}
\end{center}
\end{table}

\vspace*{-3em}
\begin{table}
\begin{center}
\parbox{0.8\linewidth}{\caption{\label{tab_fpi_mq_phys}\em
 Physical PCAC quark mass $\amqp$ and pion
 decay constant $af_\pi$ obtained from Eqs.~(\ref{eq35}) and
 (\ref{eq38}), respectively.
 The last two columns show $\amcpb \equiv \cos(\omega)\,\amqp$
 and the unrenormalised pion decay constant calculated with the
 local current $af_{v\pi}$, respectively.}}
\end{center}
\begin{center}
\begin{tabular}{|l|l|l|l|l|l|l|}
\hline
\multicolumn{1}{|c|}{$\beta$} &
\multicolumn{1}{|c|}{$a\mu$} &
\multicolumn{1}{|c|}{$\kappa$} &
\multicolumn{1}{|c|}{$\amqp$} &
\multicolumn{1}{|c|}{$af_\pi$} &
\multicolumn{1}{|c|}{$\amcpb$} &
\multicolumn{1}{|c|}{$af_{v\pi}$}
\\ \hline
0.67 & $1.0\, \cdot 10^{-2}$ & 0.1650  &0.03652(53)  &0.1672(25) & 0.03511(54) & 0.2936(63)  \\
\hline
0.67 & $1.0\, \cdot 10^{-2}$ & 0.1655  &0.02739(55)  &0.1541(25) & 0.02549(59) & 0.2750(73)  \\
\hline
0.67 & $1.0\, \cdot 10^{-2}$ & 0.1660  &0.02006(59)  &0.1447(23) & 0.01739(69) &  0.2549(84)  \\
\hline
0.67 & $1.0\, \cdot 10^{-2}$ & 0.1665  &0.01154(11)  &0.1192(18) & 0.00575(71) &  0.2113(62) \\
\hline
0.67 & $1.0\, \cdot 10^{-2}$ & 0.1670  &0.01117(38)  &0.1085(37) & -0.00497(43) & 0.1932(80) \\
\hline
0.67 & $1.0\, \cdot 10^{-2}$ & 0.1675  &0.01810(69)  &0.1203(44) & -0.01508(82) & 0.219(13) \\
\hline
0.67 & $1.0\, \cdot 10^{-2}$ & 0.1680  &0.0230(17)   &0.1146(95) & -0.0207(18) &  0.202(14) \\
\hline
\hline
0.74 & $7.5\, \cdot 10^{-3}$ & 0.1580  &0.02262(45)  &0.1170(25) & 0.02133(66) & 0.1833(57)  \\
\hline
0.74 & $7.5\, \cdot 10^{-3}$ & 0.1585  &0.01297(44)  &0.0999(26) & 0.01057(54) & 0.1625(83)  \\
\hline
0.74 & $7.5\, \cdot 10^{-3}$ & 0.1590  &0.007611(38) &0.0874(15) & -0.00129(22) & 0.1400(56)  \\
\hline
0.74 & $7.5\, \cdot 10^{-3}$ & 0.1595  &0.01245(61)  &0.0867(39) & -0.00992(78) & 0.137(10)  \\
\hline
\end{tabular}
\end{center}
\end{table}

\begin{table}
\centering
\begin{center}
\parbox{0.8\linewidth}{\caption{\label{tab_chiral_extra}\em
 Chiral extrapolation ($\mqp=0$) of the Sommer scale parameter $r_0$ and
 pion decay constant $f_\pi$.
 (This latter is denoted by $f_0 \equiv \lim_{\mqp=0} f_\pi$.)
 The scale independent combination $f_0 r_0$ is also reported.
 Only data with positive twisted quark masses have been used for the
 extrapolations, with the exception of the point at  $a\mu=0.0075$
 and $\kappa=0.1590$ which is almost at full twist.
}}
\end{center}
\begin{center}
 \begin{tabular}{|l|l|l|l|l|l|l|l|}\hline
     \multicolumn{1}{|c|}{$\beta$} &
     \multicolumn{1}{|c|}{$a\mu$} & 
     \multicolumn{1}{|c|}{$r_0/a$} & 
     \multicolumn{1}{|c|}{$a$ [fm]} & 
     \multicolumn{1}{|c|}{$a\, f_0$}&     
     \multicolumn{1}{|c|}{$f_0\,r_0$} 
    \\
\hline
0.67 & $1.0\, \cdot 10^{-2}$ & 2.845(66) & 0.1757(41) & 0.1171(59) & 0.333(10)  \\
0.74 & $7.5\, \cdot 10^{-3}$ & 3.77(20)  & 0.1326(70) & 0.0726(25) & 0.274(20)  \\
\hline
\end{tabular}
\end{center}
\end{table}

\clearpage
\begin{table}[tb]
\begin{center}
\parbox{0.8\linewidth}{\caption{\label{fit_resu_dbw}\em
 Results of the ChPT fits with DBW2 gauge action.
 Upper part: fit with both positive and negative values of $\amcp$.
 Lower part: fit with only positive values of $\amcp$.
}}
\end{center}
\begin{center}
\begin{tabular}{|c||c|c|c||c|c|c|}
\hline
 & \multicolumn{3}{|c||}{input $Z_A$} & \multicolumn{3}{|c|}{fitted $Z_A$}
\\ \hline
                                &
 $\beta = 0.67$                 &
 $\beta = 0.74$& both $\beta$   &
 $\beta = 0.67$& $\beta = 0.74$& both $\beta$
\\ \hline\hline
$Z_A(0.67)$ & $0.952(30)$ & -           &
 $0.952(30)$ & $0.8658(90)$ & -           & $0.852(14)$
\\ \hline
$Z_A(0.74)$ & -          & $0.944(74)$ &
 $0.944(74)$ & -            & $0.868(18)$ & $0.909(31)$
\\ \hline
$F_0$ [MeV] & $80.7(3.6)$ & $68.6(5.2)$ &
 $73.7(4.8)$ & $78.9(3.2)$  & $66.0(4.4)$ & $72.0(3.0)$
\\ \hline
$B(0.67)$ [GeV] & $3.20(13)$ & -      &
 $3.20(12)$  & $3.09(10)$   & -           & $3.063(94)$
\\ \hline
$B(0.74)$ [GeV] & -         & $3.31(30)$ &
 $3.16(38)$ & -         & $3.12(19)$  & $3.18(15)$
\\ \hline
$L_{54} \cdot 10^3$ & $0.98(26)$ & $0.96(26)$ &
 $1.17(28)$ & $0.50(15)$ & $0.80(23)$ & $0.74(12)$
\\ \hline
$L_{86} \cdot 10^3$ & $0.78(13)$ & $0.81(11)$ &
 $0.94(14)$ & $0.554(84)$ & $0.76(10)$ & $0.709(61)$
\\ \hline
$W_0\cdot W\cdot 10^{-3}$ [MeV$^3$] & $50(15)$ & $-21(16)$  &
 $18(17)$   & $35(12)$    & $-30(14)$  & $6.6(8.0)$
\\ \hline
$W_0\cdot \tilde{W}\cdot 10^{-3}$ [MeV$^3$] & $89(19)$ & $21(38)$ &
 $64(29)$ & $62(14)$ & $-9(22)$  & $35(12)$
\\ \hline
$\Lambda_3~/~F_0$ & $6.1(2.8)$ & $5.5(2.3)$ &
 $5.1(2.5)$ & $5.9(1.7)$ & $5.0(2.0)$ & $5.3(1.1)$
\\ \hline
$\Lambda_4~/~F_0$ & $17.1(1.4)$ & $17.0(1.4)$ &
 $18.2(1.6)$ & $14.74(69)$ & $16.2(1.1)$ & $16.86(59)$
\\ \hline
$\mathcal{L}_{min}/d.o.f.$     & $12.8(3.5)$    &$12.3(4.9)$     & 
 $13.1(7.2)$     & $9.2(1.6)$      & $11.6(2.4)$    & $9.4(1.6)$
\\
\hline\hline
$Z_A(0.67)$ & $0.952(30)$ & -           &
 $0.952(30)$ & $0.888(10)$ & -           & $0.896(11)$
\\ \hline
$Z_A$(0.74) & -          & $0.944(74)$ &
 $0.944(74)$ & -           & $0.910(18)$ & $0.880(23)$
\\ \hline
$F_0$ [MeV] & $80.3(3.4)$ & $91.2(5.4)$ &
 $83.9(4.4)$ & $79.3(3.4)$ & $89.9(4.2)$ & $82.2(2.6)$
\\ \hline
$B(0.67)$ [GeV] & $2.92(11)$ & -      &
 $2.95(11)$  & $2.85(10)$  & -           & $2.864(84)$
\\ \hline
$B(0.74)$ [GeV] & -         & $3.46(22)$ &
 $3.52(38)$ & -        & $3.39(15)$  & $3.39(11)$
\\ \hline
$L_{54}\cdot 10^3$ & $1.39(33)$ & $1.04(53)$ &
 $1.32(28)$ & $0.86(17)$ & $0.82(23)$ & $0.80(13)$
\\ \hline
$L_{86}\cdot 10^3$ & $0.92(16)$ & $0.71(20)$ &
 $0.81(15)$ & $0.70(11)$ & $0.64(13)$ & $0.649(77)$
\\ \hline
$\Lambda_3~/~F_0$ & $7.1(4.1)$ & $7.7(6.4)$ &
 $7.7(4.0)$ & $6.4(2.2)$ & $6.9(3.0)$ & $6.7(1.7)$
\\ \hline
$\Lambda_4~/~F_0$ & $19.5(2.0)$ & $17.4(2.9)$ &
 $18.6(1.6)$ & $16.47(88)$ & $16.3(1.2)$ & $16.19(69)$
\\ \hline
$\mathcal{L}_{min}/d.o.f.$& $10.1(4.9)$        &$2.7(6.5)$     & 
$5.8(7.9)$     & $7.0(2.0)$      & $2.6(2.4)$    & $4.5(1.6)$
\\ \hline
\end{tabular}
\end{center}
\end{table}

\clearpage
\begin{table}[tb]
\begin{center}
\parbox{0.8\linewidth}{\caption{\label{fit_resu_plq}\em
 Results of the ChPT fits with plaquette gauge action.
 The columns correspond to different definitions of the currents for
 $af_\pi$ and $\amcp$.
 For the definitions see Section~\ref{sec3.3}.
 Upper part: fit with both positive and negative values of $\amcp$.
 Lower part: fit with only positive values of $\amcp$.
}}
\end{center}
\begin{center}
\begin{tabular}{|c|c|c|c|c|}
\hline
& $f_{v\pi}\;\&\;\mcpb$ & $f_\pi\;\&\;\mcpb$ & $f_{v\pi}\;\&\;\mcp$ & $f_\pi\;\&\;\mcp$
\\ \hline\hline
$B$ [GeV]                               & 5.05   & 5.04   & 5.00   & 4.90
\\ \hline
$F_0$ [MeV]                             & 104.9  & 104.2  & 88.3   & 86.6
\\ \hline
$L_{86}\cdot 10^3$                        & 0.916  & 0.950  & 1.829  & 1.943
\\ \hline
$L_{54}\cdot 10^3$                        & 1.637  & 1.709  & 2.850  & 3.027
\\ \hline
$W_0\cdot W\cdot 10^{-3}$ [MeV$^3$]          & 31.5   & 28.5   & 2.9    & 6.6
\\ \hline
$W_0\cdot \tilde{W}\cdot 10^{-3}$ [MeV$^3$]  & 43.2   & 39.7   &-3.6    &-1.3
\\ \hline
$\Lambda_3/F_0$                           & 9.8    & 9.9    & 4.5    & 4.2
\\ \hline
$\Lambda_4/F_0$                           & 21.1   & 21.6   & 30.9   & 32.7
\\ \hline
$(\sum dev^2/{\sigma}^2)/{d.o.f.}$        & 2.08   & 2.19   & 4.25   & 4.16
\\ 
\hline\hline
$B$ [GeV]                                 & 5.05   & 4.33   & 4.43   & 3.95
\\ \hline
$F_0$ [MeV]                               & 98.5   & 93.9   & 90.5   & 85.8
\\ \hline
$L_{86}\cdot 10^3$                        & 0.892  & 1.466  & 1.135  & 1.836
\\ \hline
$L_{54}\cdot 10^3$                        & 1.848  & 2.705  & 2.099  & 3.155
\\ \hline
$\Lambda_3/F_0$                           & 13.6   & 9.4    & 10.1   & 6.5
\\ \hline
$\Lambda_4/F_0$                           & 22.5   & 29.5   & 24.4   & 34.0
\\ \hline
$(\sum dev^2/{\sigma}^2)/{d.o.f.}$        & 1.36   & 2.24   & 1.26   & 1.77
\\ \hline
\end{tabular}
\end{center}
\end{table}

\clearpage

\clearpage

\clearpage

\clearpage


\end{document}